\documentclass[aps,pra,twocolumn,groupedaddress]{revtex4-2}

\usepackage[a4paper, total={6in, 8in}]{geometry}

\usepackage{dcolumn}
\usepackage[colorlinks=true]{hyperref}
\usepackage{siunitx}
\usepackage{graphicx}
\usepackage{amsmath}

\begin{document}

\title{Inhibited radiative decay enhances single-photon emitters}

\author{Florian Burger}
\author{Stephan Rinner}
\author{Andreas Gritsch}
\author{Kilian Sandholzer}
\author{Andreas Reiserer}
\email{andreas.reiserer@tum.de}
 
\affiliation{Technical University of Munich, TUM School of Natural Sciences, Physics Department and Munich Center for Quantum Science and Technology (MCQST), James-Franck-Stra{\ss}e 1, 85748 Garching, Germany}
\affiliation{TUM Center for Quantum Engineering (ZQE), Am Coulombwall 3A, 85748 Garching, Germany}
\affiliation{Max Planck Institute of Quantum Optics, Quantum Networks Group, Hans-Kopfermann-Stra{\ss}e 1, 85748 Garching, Germany}
\date{\today}

\begin{abstract}
Quantum networks and modular quantum computers require efficient spin-photon interfaces, often realized using optical resonators that enhance radiative decay on a desired transition. However, this requires small mode volumes and high quality factors, which limits multiplexing capacity and demands precise frequency tuning. Here, we demonstrate an alternative approach that circumvents these bottlenecks for upscaling. Using a W1 silicon photonic crystal waveguide with a tailored photonic bandgap, we selectively inhibit unwanted decay pathways, thereby redirecting emission to the desired transition. This enables efficient photon collection over a large frequency range, allowing the resolution and individual addressing of tens of erbium dopants. Their lifetimes are preserved, or even increased, compared to bulk material. The extended mode volume of the devices enables the use of lower dopant concentrations, thereby improving emitter coherence. Our approach can be combined with Purcell enhancement and applied to other spin-qubit platforms, opening intriguing perspectives for photonic quantum technologies.
\end{abstract}

\maketitle

\section{Introduction}

\begin{figure}
	\includegraphics[scale=0.79]{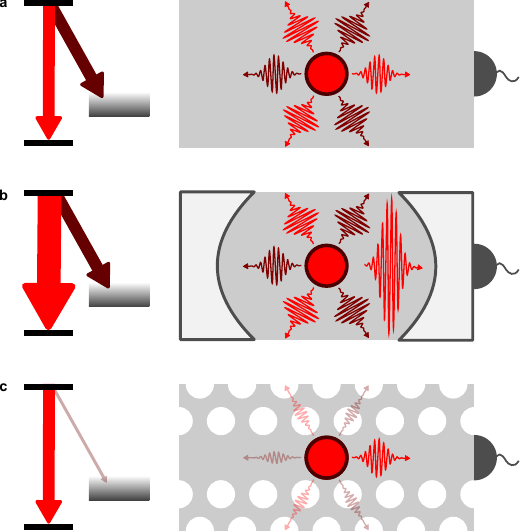}
	\caption{\textbf{Different approaches to interfacing single-photon emitters.} \textbf{a,}~~An emitter (red filled circle) in a homogeneous medium (grey) emits photons (colored wave packets) in many directions and on desired (light red) and undesired (dark red) optical transitions (arrows in the level scheme on the left). Thus, only a small fraction of the photons is emitted on the desired transition, collected and detected, e.g., with a fiber-coupled single-photon detector (black symbols on the right). This hampers the performance as a single photon source. \textbf{b,}~~ Integrating the emitter into a suited optical resonator (indicated by the two mirrors) can strongly enhance the desired transition (large red wave packet) via the Purcell effect. However, the emission into other spatial modes and on other transitions is largely unaffected. Therefore, such single photon sources require strong Purcell enhancement factors and tight light confinement to be efficient. \textbf{c,}~~In contrast, the emission into all but one spatial and spectral mode can be strongly suppressed, e.g., by photonic crystal waveguides (grey). Such devices can have an increased bandwidth and do not require precise frequency tuning and stabilization. They are not restricted to small mode volumes and thus their spectral multiplexing capacity is not limited by emitter interactions.}
    \label{fig:figure1}
\end{figure}

Controlling the generation of single photons by solid-state emitters~\cite{aharonovich_solid-state_2016} is key to scaling up photonic quantum technologies. In an ideal system, all photons should be emitted into a single mode of space, time and frequency. However, all physical systems explored so far---including trapped atoms, color centers, quantum dots, molecules, and rare-earth dopants---exhibit undesired optical decay channels that deteriorate their performance as single-photon sources. Depending on the platform, these unwanted transitions can involve phonons, other states of multilevel emitters, and spatial modes that cannot be collected efficiently (Fig.~\ref{fig:figure1}~a).

In many previous experiments, this limitation has been overcome by embedding the emitters into optical resonators that enhance the wanted transition via the Purcell effect~\cite{reiserer_colloquium_2022} (Fig.~\ref{fig:figure1}~b). However, this approach also has several downsides: First, achieving a sufficient Purcell enhancement factor \(P\) requires resonators with a small mode volume. Thus, dipolar interactions may hamper the multiplexing of many emitters in a single device. In addition, high \(P\) is only achieved in resonators with high quality factors, which restricts the bandwidth and requires sophisticated resonator fabrication and tuning techniques. Finally, high values of \(P\) can lead to very short emitter lifetimes, which is a double-edged sword. On the one hand, the resulting linewidth broadening can be beneficial for overcoming the challenges posed by spectral instabilities of the emitters~\cite{reiserer_colloquium_2022}. In addition, in short-distance connections, short lifetimes may enable a speed-up, in particular when using rare-earth dopants. On the other hand, long-distance scenarios are limited by the two-way signaling time, which, e.g., approaches $\si{\milli\second}$ timescales for typical distances of \SI{100}{\kilo\meter}. In this situation, lifetime reductions are typically not required to increase the success rate but may unnecessarily increase the demands on the experimental control and detection system while restricting spectral multiplexing capacity~\cite{chen_parallel_2020}. Also, for fast emitters, further reducing the lifetime may increase two-photon emissions in resonant excitation~\cite{fischer_signatures_2017}, which hampers most applications in distributed quantum information processing.

In this study, we therefore explore an alternative concept. Instead of enhancing a desired transition using a resonator, we use the concept of inhibited spontaneous emission~\cite{kleppner_inhibited_1981, yablonovitch_inhibited_1987} to suppress the unwanted radiative transitions (Fig.~\ref{fig:figure1}~c). The underlying approach has been studied previously in different solid-state systems. It has been shown that the lifetime~\cite{polman_erbium_2001} and the emission spectrum~\cite{zhao_suppression_2012} of emitters in nanocrystals depend on their size and surrounding refractive index. Similarly, the ratio of electric and magnetic dipole emission in rare-earth doped thin films in proximity to a gold mirror has been investigated~\cite{karaveli_spectral_2011}. Additionally, integrating solid-state emitters into nanostructured environments~\cite{yablonovitch_inhibited_1987} has enabled advanced control over the optical local density of states (LDOS)~\cite{lodahl_interfacing_2015} in waveguides~\cite{bleuse_inhibition_2011} and photonic crystals~\cite{viasnoff-schwoob_spontaneous_2005, wang_mapping_2011}. This has allowed enhancing the emitter decay into a single spatial mode while reducing the radiative lifetime, achieving near-unity beta factors with quantum dots~\cite{arcari_near-unity_2014} and color centers in diamond~\cite{ding_purcell-enhanced_2025}.

Here, instead of tailoring the LDOS such that the emission is increased, we explore the potential of inhibiting the spontaneous emission on undesired radiative transitions. Specifically, we study erbium emitters in silicon photonic-crystal waveguides (PCWs). By reducing the LDOS over a broad spectral range, we implement an efficient, broadband, and multiplexed single-photon source in the telecommunications C band in which the emitter lifetime is not reduced but preserved or even increased. As the studied effect is largely independent of the waveguide length, our approach enables multiplexing many emitters at low concentration, potentially overcoming concentration-dependent limitations on the spectral stability of erbium dopants~\cite{gritsch_purcell_2023, fruh_spectral_2026}.

Our experiment is based on erbium dopants in nanophotonic silicon waveguides with a large band gap that reduces the optical density of states over a broad spectral range. Erbium-doped silicon has recently emerged~\cite{weiss_erbium_2021} as a promising hardware platform for quantum networks that combines coherent optical transitions~\cite{gritsch_narrow_2022, berkman_observing_2023} in the minimal-loss band of optical fibers~\cite{holewa_solid-state_2025} with millisecond ground-state spin coherence~\cite{berkman_long_2025}. Furthermore, it is fully compatible with foundry-based semiconductor manufacturing~\cite{rinner_erbium_2023}. When integrated into optical resonators, the lifetime of the optical transitions can be strongly reduced~\cite{gritsch_purcell_2023}, and individual spins can be controlled and read out in a single shot~\cite{gritsch_optical_2025}. However, the small volume and narrow bandwidth of the resonators used in these experiments have restricted the spectral multiplexing capacity to a few dopants. This limitation is overcome in our current approach. We can thus spectrally resolve, individually address, and study tens of single emitters in the same device.

\section{Results}

\subsection{Photonic crystal waveguide design}

\begin{figure*}
	\includegraphics[width=\linewidth]{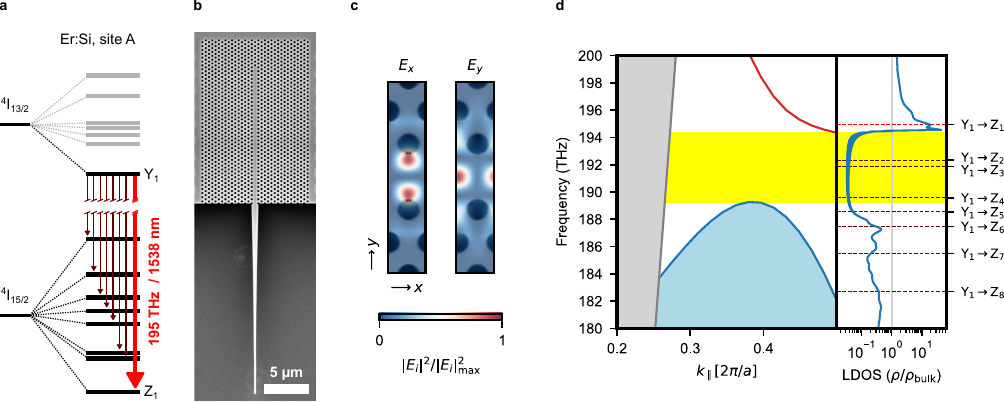}
	\caption{\textbf{Photonic crystal waveguide (PCW) for tailoring the radiative decay of individual erbium dopants in silicon.} \textbf{a,}~~Level scheme. The interaction with the host crystal splits the spin-orbit coupled energy levels \(^4\text{I}_{15/2}\) and \(^4\text{I}_{13/2}\) of erbium's 4f electrons into the crystal field levels \(Z_1\) to \(Z_8\) and \(Y_1\) to \(Y_7\), respectively. At cryogenic temperatures, the \(Y_1\) and \(Z_1\) levels cannot decay via phonon emission and thus exhibit long optical coherence. \textbf{b,}~~W1 waveguide. The scanning electron microscope image shows a representative PCW terminated by a tapered fiber coupler at the bottom and a photonic crystal mirror at the top. \textbf{c,}~~Normalized absolute value squared of the strongest electric field components, \(E_x\) (left panel) and \(E_y\) (right), of the guided eigenmode in a unit cell of the W1 waveguide. \textbf{d,}~~Simulated photonic band structure (left) and LDOS (right) for a \(y\)-dipole at the maximum field position of $E_y$. Above the line where $\omega/k_{\parallel}$ equals the speed of light (grey area), the spectrum of states is continuous. Below, a photonic band gap is formed, surrounded by slab modes (light blue) with reduced LDOS. Along the \(\Gamma-K\) direction ($k_{\parallel}$), two modes are guided in the W1 waveguide. Below the lower, even mode (red), an approximately \SI{5}{\tera\hertz} wide gap (yellow) remains. Thus, the LDOS is significantly reduced for all transitions (\(Y_1\rightarrow Z_2...Z_8\), right axis) of embedded erbium dopants, suppressing their spontaneous emission. In contrast, on the $\text{Y}_1\leftrightarrow \text{Z}_1$ transition, the slow light effect leads to an increased LDOS and an enhanced decay into the guided mode.}
	\label{fig:figure2}
\end{figure*}

In this work, erbium is integrated into silicon at lattice site ``A''~\cite{holzapfel_characterization_2025}, for which the transitions between the lowest crystal field (CF) levels \(Y_1\) and \(Z_1\) of the \(I_{15/2}\) and \(I_{13/2}\) manifolds occur within the telecommunications C band at approximately \SI{1538}{\nano\meter} (Fig. \ref{fig:figure2}~a). In bulk silicon, this transition has a small branching fraction of \SI{23\pm5}{\percent}~\cite{gritsch_narrow_2022}, meaning that the \(Y_1\) state decays predominantly to the higher CF levels with wavelengths between \(1550\) and \SI{1650}{\nano\meter}. This work aims to eliminate these unwanted decay channels, so that the excited state decays predominantly via the \(Y_1 \leftrightarrow Z_1\) transition into a single waveguide mode efficiently coupled to an optical fiber.

For that purpose, we use PCWs that feature a large photonic band gap (PBG) and a strongly dispersive guided mode. The starting point is a photonic-crystal slab, a quasi-2D photonic crystal, consisting of a triangular lattice of cylindrical air holes in an undercut, high refractive-index membrane of sub-wavelength thickness. An SEM image of such a device is shown in Fig. \ref{fig:figure2}~b. Leaving out a single row of air holes in the \(\Gamma -K\) direction creates a W1 waveguide~\cite{joannopoulos_photonic_2011} that features two guided modes, one with even (shown in Fig. \ref{fig:figure2}~c) and one with odd symmetry. The surrounding photonic crystal slab has a $\SI{50}{\tera\hertz}$ wide PBG for TE-like modes. Fig.~\ref{fig:figure2}~b shows the photonic band structure for a PCW of infinite length, which we compute using a frequency-domain eigensolver~\cite{johnson_block-iterative_2001}. The devices studied experimentally exhibit the same cross-section but different finite lengths (see methods). The PBG in between the slab modes (light blue) extends approximately from \SI{190}{\tera\hertz} to \SI{240}{\tera\hertz}. The even mode (red) at lower frequencies features a flat dispersion relation at the edge of the Brillouin zone, indicating ``slow light'', i.e., propagation with a strongly reduced group velocity~\cite{vlasov_active_2005, baba_slow_2008}. In the region between approximately \SI{190}{\tera\hertz} and \SI{195}{\tera\hertz}, a TE band gap remains (yellow), suppressing all emission at these frequencies.

Alongside the band structure, we show the normalized LDOS in a symmetric PCW that is open at both ends. It was simulated using a finite-difference time-domain method~\cite{oskooi_meep_2010} for an emitter with an in-plane electric dipole perpendicular to the waveguide direction (\(y\) dipole) at the field maximum of the guided mode (cf. Fig.~\ref{fig:figure2}~c). The LDOS shows a strong spectral dependence~\cite{javadi_numerical_2018}: In the frequency range of the even guided mode (red in Fig.~\ref{fig:figure2}~d), it is equal to or greater than one (gray line), and it increases towards the edge of the band gap (\(\rho/\rho_\mathrm{bulk} > 10\)) as the group velocity tends to zero~\cite{lodahl_interfacing_2015, gonzalez-tudela_lightmatter_2024}. Thus, the radiative decay of transitions in this frequency range is enhanced via the Purcell effect. In practice, the maximum achievable Purcell enhancement is limited by finite size effects~\cite{faggiani_implementing_2017} and ultimately by fabrication imperfections~\cite{hughes_extrinsic_2005}. In the band gap, between approximately \SI{190}{\tera\hertz} and \SI{195}{\tera\hertz}, emission is strongly suppressed (\(\rho/\rho_\mathrm{bulk} < 0.1\)). At lower frequencies, where only unguided slab modes are available, the suppression is weaker but still considerable (\(0.1 < \rho/\rho_\mathrm{bulk} < 1\)).

In addition to the spectral dependence, the LDOS exhibits a spatial variation: The Purcell enhancement is most substantial at the electric-field maximum and vanishes at positions with a small field. Fig.~\ref{fig:figure2}~d only shows the LDOS at the maximum of the electric field component $E_y$ for a dipole oriented parallel to the field; the spectra for other dipole orientations and positions in the investigated structures are presented in Sec. I of the supplementary information. There, one can see that independent of the specific position and orientation in the waveguide, the emission of erbium dopants on all unwanted transitions from the optically excited state $Y_1$ to the higher-lying crystal-field levels of the ground-state manifold ($Z_2$ to $Z_8$) will be strongly suppressed. This inhibition of spontaneous emission leads to an increase in the lifetime of the $Y_1$ state.

The effect is counteracted for dopants at the field maximum whose dipole matches the electric field orientation, such that the emission on the $Y_1\rightarrow Z_1$ transition is enhanced. By choosing suitable parameters for the geometry of the photonic crystal waveguide, the enhancement of the $Y_1\rightarrow Z_1$ transition can be precisely adjusted to compensate for the inhibition of the other transitions. In this way, the emission can be channeled almost entirely into the desired mode without changing the lifetime of the emitters.

\subsection{Spectral properties of erbium dopants in a silicon photonic-crystal waveguide}
\begin{figure*}
	\includegraphics[width=\linewidth]{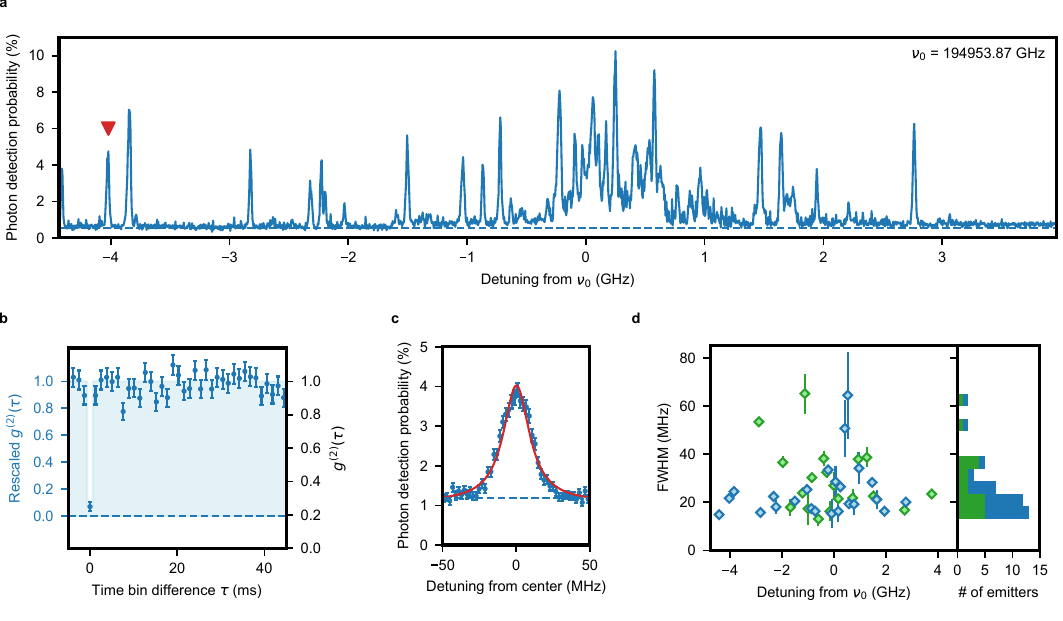}
	\caption{\textbf{Spectral properties of erbium dopants in a silicon photonic-crystal waveguide.} \textbf{a,}~~After pulsed resonant excitation around the $Y_1 \rightarrow Z_1$ transition frequency \(\nu_0\) of site ``A'', the fluorescence spectrum of an inhomogeneously broadened ensemble of erbium dopants in PCW A exhibits distinct peaks. The dashed line indicates the dark count level. \textbf{b,}~~A measurement of the autocorrelation function $g^{(2)}(\tau)$ on the same emission line results in a zero-delay value of $0.07 \pm 0.01$ when subtracting~\cite{becher_nonclassical_2001} the contribution of the independently measured detector dark counts (left axis), or $0.25\pm0.03$ without correction (right axis). This proves that the peak originates from a single dopant. \textbf{c,} ~~A Lorentzian (red) fitted to a high-resolution measurement of a single emission line (marked by the red triangle in a) gives a linewidth of \SI{21.5\pm0.1}{\mega\hertz}. Error bars: 1 S.D. \textbf{d,}~~The spectral diffusion linewidth measured on PCWs A and B (blue and green, respectively) slightly differ between the emitters. It exceeds \SI{13}{\mega\hertz} for all dopants. The average linewidth is \SI{27}{\mega\hertz} with a standard deviation of \SI{12}{\mega\hertz}. The error bars indicate the standard error of the Lorentzian fits.}
	\label{fig:fig3_panorama_g2_LW}
\end{figure*}

To demonstrate the described tailoring of the radiative decay, we perform pulsed resonant fluorescence spectroscopy~\cite{gritsch_purcell_2023} on three different PCWs on the same chip, labeled A, B, and C, that have identical design parameters except for their different lengths (see methods). The devices are mounted in a closed-cycle cryostat at \SI{1.2}{\kelvin}. The dopants are repeatedly excited by laser pulses of $\SI{2}{\micro\second}$ duration while the excitation laser frequency is swept. After each pulse, the fluorescence is detected by a single-photon counter. Averaging over 8000 repetitions at each frequency setting, we obtain the fluorescence spectrum in Fig. \ref{fig:fig3_panorama_g2_LW}~a. The implantation dose of \(\SI{1e11}{\per\centi\meter\squared}\) was chosen such that we expect ~40 emitters in each waveguide when assuming an integration yield of \SI{1}{\percent} in site A~\cite{gritsch_narrow_2022}. Approximately thirty spectrally separated peaks are observed around the corresponding optical transition frequency. 

To demonstrate that the peaks originate from single dopants, we measure the autocorrelation function of several of the isolated lines and consistently obtain values of \(g^{(2)}(0)<0.5\), see e.g. Fig. \ref{fig:fig3_panorama_g2_LW}~b. In a high-resolution measurement, Fig. \ref{fig:fig3_panorama_g2_LW}~c, the emitters exhibit Lorentzian line shapes (red fit curve), approximately four orders of magnitude broader than the lifetime limit. Fig. \ref{fig:fig3_panorama_g2_LW}~d shows the extracted linewidth of all peaks; the individual fits are displayed in the supplementary information. The width slightly differs between the emitters, with a minimum value of approximately $\SI{13}{\mega\hertz}$ that is comparable to previous results in nanophotonic resonators~\cite{gritsch_purcell_2023, gritsch_optical_2025}. As the homogeneous linewidth in Er:Si is much narrower, down to 10 kHz at the temperature of the present experiment~\cite{gritsch_narrow_2022}, the broadening is attributed to spectral diffusion that originates from charge or spin noise in the devices~\cite{wolfowicz_quantum_2021}.

\subsection{Optical lifetime}
\begin{figure*}
	\includegraphics[width=\linewidth]{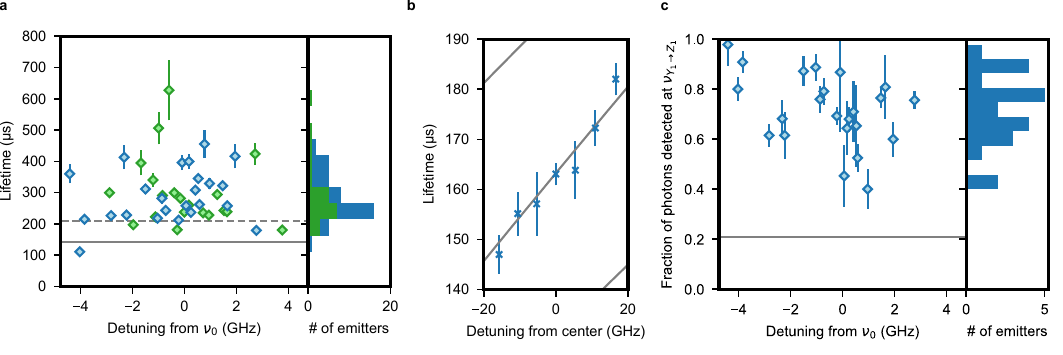}
	\caption{\textbf{Optical lifetimes and \(Y_1\rightarrow Z_1\) emission of single dopants.} \textbf{a,}~~Exponential fits are used to extract the lifetimes of all individual emitters in the inhomogeneous ensembles of the PCWs A and B (blue and green, respectively), shown as a function of the detuning from the center of the inhomogeneous line. The solid gray line marks the lifetime in bulk silicon of \SI{142\pm1}{\micro\second} while the dashed line denotes the value in silicon strip waveguides of \SI{209\pm1}{\micro\second}~\cite{gritsch_narrow_2022}. The fluctuation of the lifetime, further analyzed in the histogram on the right, stems from the random position and orientation of the emitters relative to the guided mode. The average lifetime is \SI{295}{\micro\second} with a standard deviation of \SI{97}{\micro\second}. \textbf{b,} A magnetic field is applied to tune the emission frequency of a single emitter in PCW C via the Zeeman effect. The spectral dependence of the local density of states (LDOS) can then be determined by extracting the lifetime. The gray curves show the expectation based on the LDOS simulation and the measured dispersion in three different PCWs with parameters similar to those of the device hosting the emitters (see supplementary information). While the slope can be accurately predicted, slight variations across the nanofabricated devices lead to a deviation of the curve offset. \textbf{c,}~~As an effect of the selective suppression of unwanted optical transitions, the fraction of the photons that are emitted on the \(Y_1\rightarrow Z_1\) transition is increased. It can thus exceed the value measured in a silicon strip waveguide of \SI{23\pm5}{\percent} (gray line). Inserting a narrowband spectral filter allows determining the fraction of light that is detected at the \(Y_1\rightarrow Z_1\) frequency (blue data points). It approaches unity for some emitters but exhibits large fluctuations (histogram on the right) because of the random position and orientation of the emitters in the guided mode of the PCW. The average is \SI{72}{\percent} with a standard deviation of \SI{14}{\percent}. The measurement was performed on PCW A. All error bars denote 1 S.D.}
	\label{fig:figure4}
\end{figure*}

After studying the spectral properties of the emitters, we now turn to the lifetime of the optically excited state. The waveguides in this study are designed such that the increase in lifetime due to the suppression of the unwanted transitions \(Y_1\rightarrow Z_2 ... Z_8\) is approximately balanced by the enhancement of the \(Y_1\rightarrow Z_1\) decay for optimally positioned and oriented dopants, leaving the lifetime close to unchanged from its bulk value of \SI{142(1)}{\micro\second}~\cite{gritsch_narrow_2022}. This enables us to explore the inhibition of spectral decay channels without compromising the signal-to-noise level, which would be limited by detector dark counts in case the emitter lifetime gets too long. While the suppression is approximately independent of the dopant position (see supplementary information), the enhancement of \(Y_1 \rightarrow Z_1\) varies significantly with the spatial intensity profile of the guided mode (cf.~\ref{fig:figure2}b). As the emitters are implanted homogeneously, their random orientations and locations in the waveguide will lead to a statistical spread in the lifetime.

To study this effect, we perform a time-resolved analysis of the fluorescence data recorded within \SI{600}{\micro\second} after each excitation pulse. Fig.~\ref{fig:figure4}~a shows the lifetimes of the $Y_1$ level of the emitters in two different waveguides on the same chip, extracted from exponential fits to the fluorescence decay. As expected, the lifetimes in the ensemble vary strongly due to the random emitter integration in the PCW. They range from around \SI{100}{\micro\second} up to \SI{600}{\micro\second}. The associated histogram shows an asymmetric distribution in which most dopants have a lifetime between \SI{200}{\micro\second} and \SI{300}{\micro\second}. A numerical modeling of the distribution would require precise knowledge of the dipolar transition operators in Er:Si, which is left for future work. A correlation between lifetimes and linewidths is neither expected nor observed.

Furthermore, within the inhomogeneous linewidth, no clear dependence on the detuning of the emitters is observed, proving the broadband nature of our approach. Our observations agree with the expectation from the LDOS simulation, which only shows a small frequency dependence within the spectral range of the inhomogeneous broadening that is obscured by the fluctuations caused by the random emitter orientation and position. However, at larger detunings of tens of \si{\giga\hertz}, one expects a significant effect of the slope of the dispersion relation on the lifetime.

To investigate this, we apply a magnetic field of up to \SI{3}{\tesla} that splits the $Y_1 \rightarrow Z_1$ transition via the Zeeman effect~\cite{holzapfel_characterization_2025}. In this way, the radiative lifetime of a single dopant can be studied as a function of its detuning from the band edge. The result is shown in Fig. \ref{fig:figure4}~b. The observed change of the lifetime with the emission frequency is consistent with a model based on the simulated DOS and the measurement of the dispersion of three different waveguides (gray curves, see supplementary information for details). In the depicted spectral range, the curve can be approximated by a straight line; this changes on larger scales and when approaching the van-Hove singularity at the edge of the bandgap~\cite{lodahl_interfacing_2015}. Thus, adjusting the detuning from the band edge allows tuning this emitter's lifetime to precisely match the bulk value or that of another dopant. This paves the way toward two-photon interference with high contrast and thus for spin-photon~\cite{uysal_spin-photon_2025} and remote spin-spin entanglement experiments \cite{ruskuc_multiplexed_2025}.

\subsection{Relative increase of the \(Y_1\rightarrow Z_1\) emission}

The extended excited state lifetimes described in the previous section already indicate that radiative transitions are efficiently suppressed as a result of the tailored LDOS of the PCW. Now, we demonstrate that this suppression is spectrally selective and that the emission is efficiently channeled into the $Y_1 \leftrightarrow Z_1$ transition.

To investigate the spectral dependence of the emission, we insert a tunable narrow-band color filter (FWHM \SI{0.1}{\nano\meter}) into the detection path and repeat the measurement. We then fit the observed peaks with Lorentzian lines and compare their amplitudes $I_\text{filter}$ to those of the corresponding peaks in the unfiltered measurement $I_0$ while correcting for the independently calibrated filter transmission $\chi$. Fig.~\ref{fig:figure4}~c shows the relative fraction of photons that are detected on the $Y_1 \rightarrow Z_1$ transition, $\frac{I_\text{filter}}{I_0 \cdot \chi}$. As the coupling of the emitted light to the waveguide depends on the position and orientation of the emitters relative to the field maxima of the guided mode, which is random in our experiment, large fluctuations are observed. In the range of the inhomogeneous distribution of emitters in Er:Si, the LDOS is almost constant (see supplement); therefore, the branching fraction shows no significant spectral dependence. For all dopants in the ensemble in PCW A, more than \SI{40}{\percent} of the detected photons exhibit the desired transition frequency. For some emitters, even values close to unity are reached. For comparison, the horizontal line indicates the branching fraction of erbium in silicon rib waveguides without a tailored LDOS, \(P_{Z1, \text{bulk}} = \SI{23\pm5}{\percent}\), measured using the same technique~\cite{gritsch_narrow_2022}.

Thus, we find that in the fiber-coupled output, the fraction of light at the desired transition frequency is increased. In principle, this may have two causes: First, the PCW or the fiber-optical setup may act as a spectral filter that transmits light at the \(Y_1\rightarrow Z_1\) transition frequency but absorbs or scatters photons at the frequencies of \(Y_1\rightarrow Z_2 ... Z_8\), thus preventing them from reaching the detectors. Second, the increase may be due to a change in the branching fraction because of a spectrally selective inhibition of the emission. To exclude that our observations are merely caused by the former effect, spectral filtering, we calculate the expected probability of detecting a photon at the \(Y_1 \rightarrow Z_1\) transition frequency when assuming that the PCW has no effect on the branching ratio, such that the excited state still decays via the \(Y_1 \rightarrow Z_1\) transition with the bulk probability of \(P_{Z1, \text{bulk}}=\SI{23\pm5}{\percent}\). The full calculation (see Sec. VII of the supplementary information) assumes that the PCW is lossless, that there is no non-radiative decay, that the emission is coupled into the waveguide with 100\,\% efficiency, and that the emitter is excited with 50\,\% efficiency, which is the maximum value possible upon incoherent excitation, as observed at the used laser power and dopant concentration~\cite{fruh_spectral_2026} (see supplementary information). Even in this best-case scenario, one would expect that a photon is detected after the filter in less than \SI{1.84\pm0.43}{\percent} of the repetitions of the experiment if the branching were unchanged. This is clearly below the experimental value of \SI{3.41\pm0.23}{\percent}. Thus, the increased fraction of light detected on the \(Y_1 \rightarrow Z_1\) transition cannot be explained by mere spectral filtering. Instead, the measurement proves that the inhibition of the radiative decay channels is spectrally selective, and that the branching fraction of the \(Y_1 \rightarrow Z_1\) transition is increased compared to its bulk value. As this transition is not broadened by subsequent phonon-induced relaxation to lower ground-state CF levels, the single-photon properties of the emitters are improved, which paves the way for quantum photonics applications.

To this end, the efficiency of the photon source will be paramount. In our current demonstration, by correcting for the independently characterized losses of the detection setup (see supplement), we can infer an efficiency of \SI{20\pm2}{\percent} in the on-chip waveguide, and \SI{14\pm1}{\percent} after coupling to a single-mode fiber, comparable to the state-of-the-art of single-photon sources at telecommunications wavelength (c.f. Table 2 in \cite{holewa_solid-state_2025}). The achieved numbers can be further improved by increasing the excitation probability (currently $<50\,\%$) using optical $\pi$ pulses or pulsed excitation to a higher CF level, and by increasing the LDOS of the \(Y_1 \rightarrow Z_1\) transition by operating closer to the band gap and/or with structures that enable even larger group indices~\cite{baba_slow_2008, faggiani_implementing_2017}.

\section{Discussion}

By suppressing the unwanted radiative decay channels of single-photon emitters, we demonstrate a novel approach to implementing efficient single-photon sources. In contrast to earlier works that used nanophotonic resonators~\cite{reiserer_colloquium_2022}, the larger spatial extent and the spectrally broad emission channeling of photons into the guided mode of PCWs allow for controlled coupling of many emitters while avoiding interactions. Thus, our approach increases the capability for spectral multiplexing, as the number of addressable emitters is only limited by spectral diffusion and by the inhomogeneous width of the emitter ensemble rather than by the spectral width or mode volume of the photonic device. The latter can be changed freely by adjusting the waveguide length without an immediate effect on the LDOS.

Our current work focuses on suppressing unwanted radiative transitions while providing only moderate Purcell enhancement of the desired one via the slow-light effect. However, in some situations, it may still be desirable to shorten the emitters' lifetimes, e.g., in order to overcome their spectral instability~\cite{fruh_spectral_2026}. In this case, the Purcell enhancement may be increased in devices with a higher group index~\cite{baba_slow_2008}, which may use optimized couplers to the slow light region~\cite{faggiani_implementing_2017}. This can enable Purcell enhancements comparable to those achieved in nanophotonic cavities~\cite{reiserer_colloquium_2022} at similar bandwidths while preserving much larger mode volumes and thus multiplexing capacity. In addition, our work may also stimulate novel designs of nanophotonic resonators that combine LDOS enhancements on resonance with broadband inhibition of undesired radiative transitions. However, our approach can also be used to engineer efficient spin-photon interfaces with increased optical lifetime. This can be advantageous in ensemble-based quantum memories that store impinging photons in an optically excited state of the emitters~\cite{craiciu_multifunctional_2021}, and in precision spectroscopy and all-optical sensing experiments whose resolution can be limited by the emitter lifetime~\cite{tiranov_sub-second_2026}.

For erbium in silicon, our approach can be combined with recent advances in spin control \cite{gritsch_optical_2025} to realize a spin-photon interface, where the spin can be read out optically after aligning the radiative dipoles with a vector magnet to increase the cyclicity~\cite{raha_optical_2020}. Using glide-plane PCWs, chiral spin-photon interfaces~\cite{lodahl_chiral_2017} can be implemented to enable, e.g., spin-dependent emission into different directions. The multiplexing capability of the devices may be further increased by suppressing spectral diffusion, which is attributed to spin and charge noise in our samples. Thus, a considerable improvement is expected when applying strong electric fields to ionize charge traps, as demonstrated earlier with quantum dots~\cite{uppu_quantum-dot-based_2021} and color centers~\cite{anderson_electrical_2019}. Still, even without these advances, frequency-multiplexed optical entanglement of remote dopants can be achieved in our devices via tailored rephasing protocols~\cite{ruskuc_multiplexed_2025}. To this end, it would be beneficial to reduce the fluctuation of the lifetimes observed in our experiments, which can be achieved by spatially selective implantation~\cite{hollenbach_wafer-scale_2022} and by engineering the band structure to obtain flat bands with constant group index over an extended spectral range~\cite{baba_slow_2008}. This would open the door to exploring the effects of collective couplings in solid-state light-matter interfaces~\cite{tiranov_collective_2023, gonzalez-tudela_lightmatter_2024}.

While our experiment used erbium dopants in silicon, it is not limited to this combination of emitter and host. Instead, our approach can be directly transferred to many other solid-state emitters, provided their lifetime is not dominated by nonradiative decay channels~\cite{wolfowicz_quantum_2021}. This includes rare-earth emitters in various host crystals~\cite{ruskuc_multiplexed_2025, ourari_indistinguishable_2023, chen_parallel_2020}, color centers in silicon~\cite{durand_broad_2021, higginbottom_optical_2022, komza_indistinguishable_2024, johnston_cavity-coupled_2024}, in silicon carbide~\cite{wolfowicz_vanadium_2020, lukin_integrated_2020, cilibrizzi_ultra-narrow_2023} and in diamond~\cite{wan_large-scale_2020}. In particular, as the bandgap of photonic crystals can span tens of $\si{\tera\hertz}$~\cite{jia_two-pattern_2011} and thus surpass the frequency of localized and propagating phonon modes, our approach can be used to fully suppress phonon sideband radiation that has limited the rates in many pioneering quantum networking experiments with solid-state emitters~\cite{hensen_loophole-free_2015}. Thus, our demonstration of inhibited spontaneous emission paves the way for the implementation of various photon sources that emit efficiently into a single mode and for robust quantum network nodes that do not require precise resonator tuning and stabilization.

\section{Methods}

\subsection{Device design}\label{w1design}
The W1 waveguides used in this work are formed by leaving out a row of holes along the \(\Gamma-K\) direction of a triangular lattice of circular air holes in a thin slab of silicon. The slab thickness is \SI{220}{\nano\meter}, the lattice constant used is \(a=\SI{420}{\nano\meter}\) and the hole radius measured by scanning electron microscopy after fabrication is \(r=0.28a\). The photonic crystals studied here all have a finite width of 33 rows and the waveguides two different lengths of 42 (PCWs A and B) and 19 (PCW C) periods. This short length avoids the complication of Anderson localization observed in longer waveguides with higher group indices~\cite{sapienza_cavity_2010}. On one end, the PCWs are terminated by reinserting 7 holes into the waveguide, which creates an efficient mirror that ensures all light is emitted into one direction. On the other end, all PCWs feature a short fast-light section with a lower group index to improve the coupling efficiency. This step coupler consists of 4 periods with a stretch factor of 1.07~\cite{hugonin_coupling_2007}. At the wavelength of interest, this leads to a high transmission of light from the slow-light section to an exponentially tapered \(\SI{20}{\micro\meter}\) long fiber coupler~\cite{tiecke_efficient_2015}. To measure the group index, we fabricate the PCWs in pairs where for one of the waveguides, the interface between the fast- and slow-light sections is made partially reflective by inserting an additional hole. In this way, a Fabry-Perot-like resonator is formed between the end mirror and the reinserted hole. By measuring its free spectral range, the effective group index of the waveguide can be determined~\cite{arcari_near-unity_2014}. A corresponding measurement of the reflection spectrum is shown in the supplementary information.

\subsection{Device fabrication}
The devices are fabricated similarly to those used in reference~\cite{gritsch_optical_2025}. The starting point is a chip diced from a commercial Czochralski-grown silicon-on-insulator wafer (SOITEC) with a \(\SI{220}{\nano\meter}\) thick device layer. Erbium is implanted by Innovion at a dose of \(\SI{1e11}{\per\centi\meter\squared}\) and an energy of \(\SI{250}{\kilo\electronvolt}\), resulting in a Gaussian depth profile, approximately centered in the device layer with a straggle of \(\SI{\approx 20}{\nano\meter}\) and a peak erbium concentration of \(\SI{1e16}{\per\centi\meter\cubed}\). The chips were first annealed at a temperature of \(\SI{500}{\celsius}\) with a \(\SI{1}{\minute}\) ramp duration starting from room temperature and a \(\SI{30}{\second}\) hold time. This protocol has previously been found to be the optimal procedure for healing implantation damage without affecting the dopants~\cite{rinner_erbium_2023}. The PCWs are then patterned using electron-beam lithography (Nanobeam Ltd., nb5) on a positive-tone resist (ZEP 520A) and subsequently transferred to the silicon device layer by reactive ion etching (Oxford Instruments PlasmaPro 100 Cobra) with fluorine chemistry at cryogenic temperatures. In a final step, the sacrificial layer of silicon oxide is partially removed with hydrofluoric acid, leaving the PCW and the taper for fiber-coupling suspended in air.

\subsection{Experimental setup}
The sample is mounted in a closed-cycle helium cryostat (ICEoxford 1K DryICE) on a three-axis nanopositioning system (Attocube ANPx312, ANPz101). Resonant fluorescence spectroscopy measurements are performed using an optical pulse setup similar to the one used in reference~\cite{gritsch_optical_2025}. The pulses are generated from a continuous-wave laser system (Toptica CTL) using two acousto-optic modulators (Gooch\&Housego Fiber-Q). In addition, a single-sideband IQ-modulator (iXblue MXIQER-LN-30) with a bias controller (iXblue MBC-IQ-LAB-A1) is used to enable frequency sweeps over the range of several GHz. 

Experimental sequences are implemented using an arbitrary waveform generator (Zurich Instruments SHFSG). For optimized dopant excitation, we use linearly chirped pulses with a length of \SI{2}{\micro\second} and a chirp-width of \SI{10}{\mega\hertz}. 
On-chip coupling is achieved using an adiabatic coupler that consists of a \SI{20}{\micro\meter} tapered waveguide contacted to a tapered optical fiber (chemically stripped, \SI{3}{\degree} taper angle). Fluorescence emitted by the dopants is separated from the excitation pulse using a beamsplitter with a splitting ratio of 95:5 (Evanescent Optics Inc.). It is detected by a superconducting nanowire single-photon detector (ID Quantique) with \SI{75\pm5}{\percent} detection efficiency. A fast optical switch (Agiltron Ultra-fast Dual Stage SM NS 1x1 Switch) is used to prevent blinding the detectors by the excitation pulse. For the filter measurement, we use an electrically tunable optical filter (WL Photonics) with a \SI{0.11}{\nano\meter} FWHM Gaussian-shaped bandwidth.

\section{Data availability}

The datasets generated and analyzed during the current study are available in the mediaTUM repository via the DOI 10.14459/2026mp1856417.

\section{Acknowledgements}

We thank Suyash Gaikwad for his technical contributions to the implementation of the group index measurement method. Nanofabrication was performed using facilities at the \textit{Walther-Meißner-Institut} and the \textit{Center for Nanotechnology and Nanomaterials (TUM)} in Garching.

\section{Funding}
Deutsche Forschungsgemeinschaft (DFG, German Research Foundation) under Germany's Excellence Strategy - EXC-2111 - 390814868 and via the individual grants 452035973 and 559594594; German Federal Ministry of Research, Technology and Space (BMFTR) via the grant agreements No 13N16921, 16KISQ046 and 16KIS2198; Munich Quantum Valley, which is supported by the Bavarian state government with funds from the Hightech Agenda Bayern Plus; European Union (ERC CoG OpeNSpinS, grant agreement 101170219). Views and opinions expressed are, however, those of the author(s) only and do not necessarily reflect those of the European Union or the European Research Council. Neither the European Union nor the granting authority can be held responsible for them.

\section{Author contributions}

F.B. performed the simulations, designed and fabricated the sample. A.G. and S.R. built the measurement setup. S.R., F.B. and K.S. carried out the measurements. F.B. and K.S. analyzed the data. F.B. and A.R. wrote the manuscript with input from all authors. A.R. supervised the project.

\section{Competing interests}

The authors declare no competing interests.

\widetext
\clearpage

\begin{center}
\textbf{\large Supplementary information: Inhibited radiative decay enhances single-photon emitters}
\end{center}
\setcounter{equation}{0}
\setcounter{figure}{0}
\setcounter{table}{0}
\setcounter{page}{1}
\makeatletter
\renewcommand{\theequation}{S\arabic{equation}}
\renewcommand{\thefigure}{S\arabic{figure}}
\renewcommand{\bibnumfmt}[1]{[S#1]}
\renewcommand{\citenumfont}[1]{S#1}

\section{Local density of states simulations}

The local density of states (LDOS) in a photonic-crystal waveguide (PCW) is spatially and spectrally anisotropic. To predict the branching ratio and lifetime of embedded erbium dopants, we simulate the frequency dependence of the LDOS for different dipole orientations and at several positions using a finite-difference time-domain (FDTD) method, specifically the \texttt{dft\_ldos} function of MEEP~\cite{oskooi_meep_2010}. It excites a dipole source at a specified point inside the simulation cell and computes the radiated power by accumulating the Fourier transforms of the electric field. In the band gap of the PCW, where emission is strongly suppressed and only little power is radiated, this method is prone to artifacts in the form of high-frequency oscillations that could only be avoided using prohibitively long simulation times. Instead, we apply a second-order Butterworth lowpass filter with a cutoff frequency of 0.2 times the Nyquist frequency to the simulated spectra to remove the artifacts.

For normalization, the filtered spectra are divided by the simulated LDOS spectrum of a dipole in bulk silicon. The LDOS spectrum in the main text (Fig.~2~d) is simulated for a \(y\) dipole at the field maximum of a PCW (cf. Fig.~2~c of the main text) with a total length of 40 periods that includes 4-period step couplers at each end, which connect the PCW to strip waveguides terminated by perfectly matched layers to fully absorb the incident light. This way, the structure exhibits an LDOS spectrum that closely resembles that of an infinitely long waveguide~\cite{faggiani_implementing_2017}.

\begin{figure}
	\centering
	\includegraphics[width=1\linewidth]{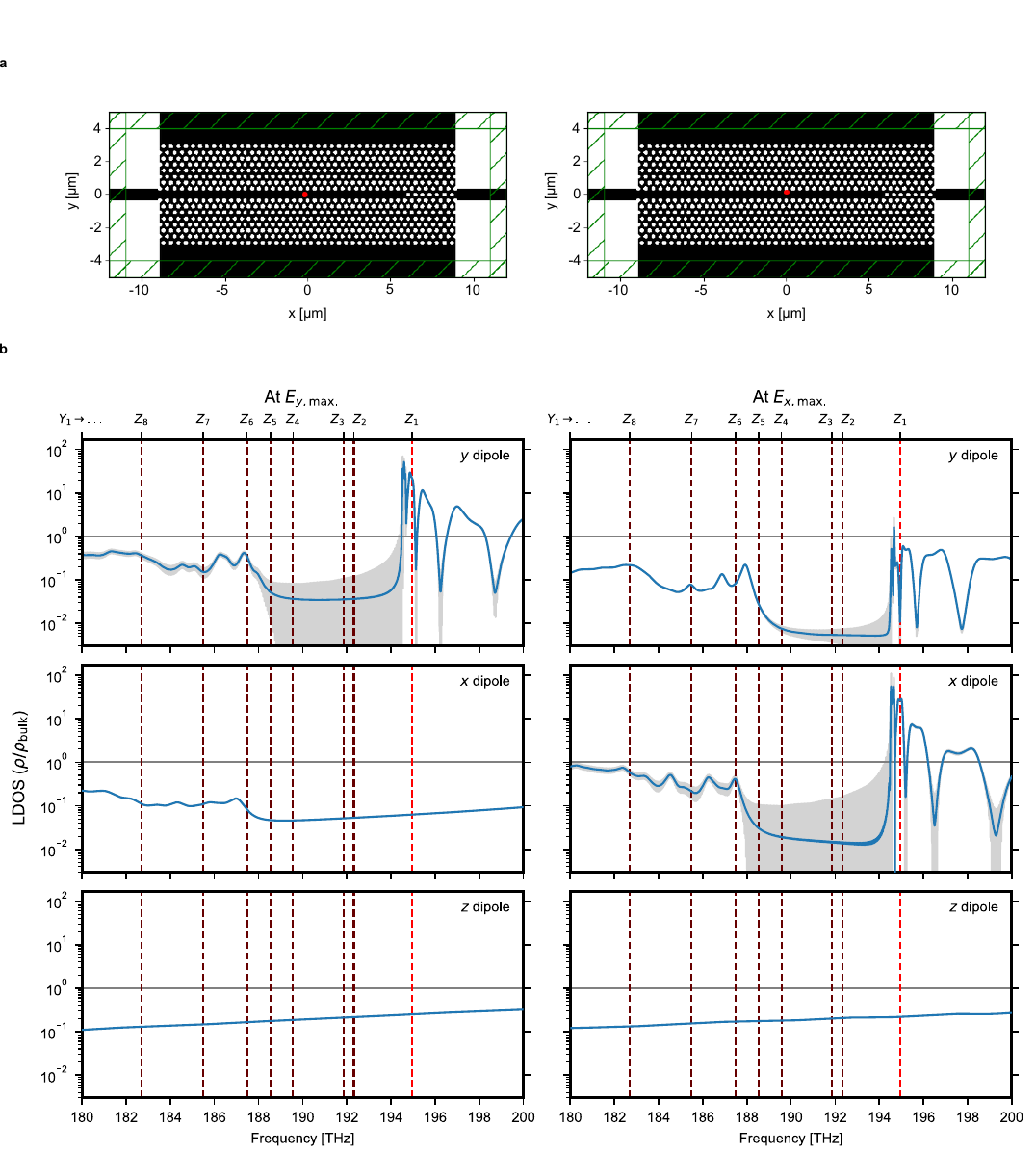}
	\caption{Local density of states (LDOS) simulation. \textbf{a}, ~PCW geometries used for simulating the LDOS spectra for different dipole orientations at the \(E_y\) (left) and \(E_x\) (right) maximum. The images show a top view of the photonic crystal slab. The black parts are silicon with a refractive index of 3.45 (at \(T<\SI{10}{\kelvin}\)), the white parts are air with a refractive index of 1. The dashed green area indicates perfectly matched layers (PML) used to create absorbing boundary conditions. The red dot marks the location of the source in each case. \textbf{b},~LDOS spectra for dipoles oriented in-plane,  either parallel to the waveguide axis (\(x\)), perpendicular to the waveguide axis (\(y\)), and out-of-plane (\(z\)). The spectra are simulated at the position of the \(E_y\) electric field component maximum or the \(E_x\) electric field component maximum of the eigenmode of the W1 waveguide, as indicated in \textbf{a}. The simulated spectra (gray) exhibit fast, high-amplitude oscillations in the band gap, where emission is strongly suppressed. These artifacts originate from the finite run time of the simulation. To remove the fast oscillations, a second-order Butterworth lowpass with a cutoff frequency of 0.2 times the Nyquist frequency filter is applied to the spectra. The filtered spectra are shown in blue. The dashed horizontal line marks \(\rho/\rho_{\mathrm{bulk}}=1\), separating the regimes of enhancement (above) and suppression (below) of the emission. The vertical red lines indicate the optical transitions into the different crystal field levels of Er:Si in site ``A''.}
	\label{fig:supplementldos}
\end{figure}

In the experiment, PCWs with a 31-period slow-light section are used. They are connected to a strip waveguide using a 4-period step coupler on one end. On the other end, the waveguide is terminated by a mirror to be able to collect all fluorescence in reflection. This leads to slightly modified spectra, which are shown in Fig.~\ref{fig:supplementldos}~b for \(x\), \(y\) and \(z\) dipoles at the maxima of \(E_y\) and \(E_x\), respectively. The simulated geometries with the position of the dipole source marked as a red dot are depicted in Fig.~\ref{fig:supplementldos}~a.

Just like in the quasi-infinite PCW, the emission of a \(y\) dipole located at the maximum of the \(E_y\) field component is enhanced for wavelengths below \(\approx\SI{1540}{\nano\meter}\), strongly inhibited between \(\approx\SI{1590}{\nano\meter}\) and \(\approx\SI{1540}{\nano\meter}\) and less strongly inhibited above \(\approx\SI{1590}{\nano\meter}\). The mirror at the end of the waveguide gives rise to a standing wave interference pattern, which leads to an oscillation of the LDOS spectrum that depends on the emitter position, see Fig.~\ref{fig:supplementldos}. The spectral position of the interference minima changes with the exact emitter position and therefore averages out for an ensemble of emitters evenly distributed across the PCW. In our measurements, we did not observe an effect of the finite waveguide length. The spectrum for an \(x\) dipole at the maximum of \(E_x\) looks very similar. Emission is suppressed across the entire spectrum for \(y\) dipoles at the \(E_x\) maximum and \(x\) dipoles at the \(E_y\) maximum. The emission of \(z\) dipoles is suppressed, independent of the position in the \(xy\) plane of the PCW.

\section{Group index measurement}

The enhancement of spontaneous emission on the $Y_1 \rightarrow Z_1$ transition is directly proportional to the group index of the devices~\cite{lodahl_interfacing_2015}, which is thus critical for the modeling and understanding of the devices. Therefore, it is essential to quantify its value and to determine the expected systematic wavelength shift of the devices compared to the simulation. To this end, we fabricate a dedicated additional waveguide next to each structure on the chip. These characterization waveguides feature an extra air hole precisely at the interface of the step coupler and the slow-light section. The effect of the additional hole is to induce a small reflection. Together with the end reflector, a Fabry-Perot resonator is formed whose free spectral range $\Delta\nu$ depends on the group index \(n_g\) and the resonator length \(L\):
\begin{equation}
	\Delta\nu = \frac{c}{2n_gL}.
\end{equation}

\begin{figure}
    \centering
    \includegraphics[width=1\linewidth]{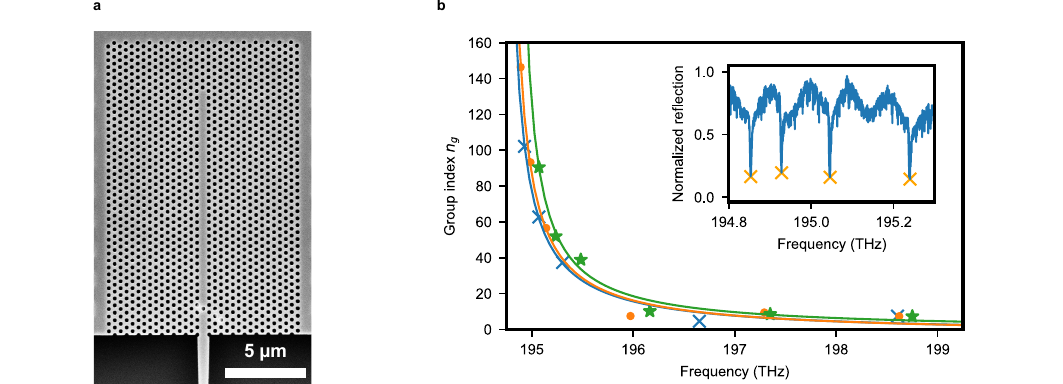}
    \caption{\textbf{a},~~Scanning electron microscope image of a PCW in a resonator configuration. A hole is reinserted between fast-light and slow-light section, making this interface partially reflective and thus forming a Fabry-Perot resonator. \textbf{b},~~The group index is inferred from the free spectral range of the Fabry-Perot resonator. To this end, the reflection spectrum of the structure is measured (inset). The separation between the observed resonances (orange crosses) is then used to determine the group index of waveguide E (main panel). Three different PCWs with identical design show similar group indices but small deviations because of fabrication tolerances: waveguide D (blue crosses), waveguide E (orange circles) and waveguide F (green stars). The data is fit with a reciprocal function (solid lines).}
    \label{fig:supplementng}
\end{figure}

By extracting $\Delta\nu$ from the reflection spectra, the group index can thus be determined. The results from three PCWs, proximal to the structures used in the main text and with nominally identical hole size and lattice constant, are shown in Fig. \ref{fig:supplementng}. The data is fit with a reciprocal function, as the group velocity increases approximately linear in frequency when approaching the band edge of a W1 waveguide~\cite{lodahl_interfacing_2015}. The measurements show that statistical fabrication offsets between the individual devices lead to a slight shift of the bandgap of the PCWs and, as a result, to considerable variations in the group index at a given frequency. The obtained \(n_g\) of up to 140 lead to a significant enhancement of the emission of the wanted $Y_1 \rightarrow Z_1$ transition that falls in the slow-light region. For an optimally oriented and positioned two-level emitter in a PCW, this enhancement is
\begin{equation}
	\label{eq:w1purcell}
	F^{\mathrm{max}}_\mathrm{P}(\omega) = \left(\frac{3}{4\pi n}\frac{\lambda^2/n^2}{V_{\mathrm{eff}}/a}\right)n_g(\omega),
\end{equation}
where \(n_g(\omega)=c/v_g(\omega)\) is the group index, \(V_{\mathrm{eff}}\approx a(\lambda/n)^2/3\) is the effective mode volume, $\lambda$ the wavelength, $n$ the refractive index, and $a$ the lattice constant~\cite{lodahl_interfacing_2015}. Thus, the decay rate increases linearly with $n_g(\omega)$ and the lifetime of embedded emitters will exhibit a frequency dependence. This effect is shown in Fig.~4~b of the main text. The details of this calculation will be described in the next section.

\section{Theoretical model for the branching contrast and optical lifetime}

\begin{figure}
	\centering
	\includegraphics[width=1\linewidth]{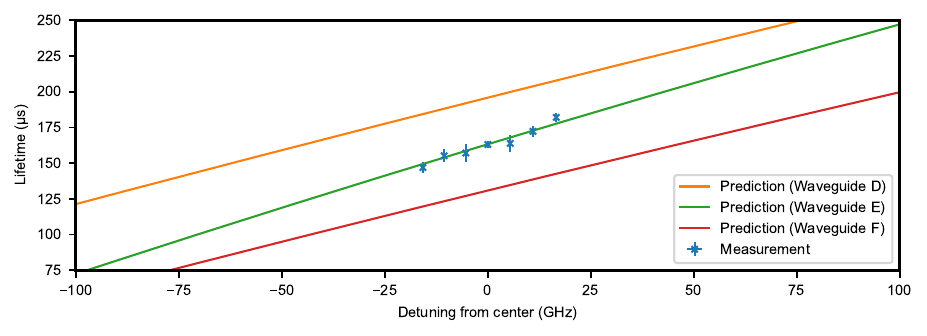}
	\caption{Frequency dependence of the optical lifetime of the \(Y_1\) level. To calculate the expected lifetime, we measure the group index \(n_g\) of three waveguides with identical design and simulate the LDOS inhibition on the \(Y_1 \rightarrow Z_2\) to \(Y_1 \rightarrow Z_8\) crystal-field transitions. This model (colored curves) agrees well with the data (blue crosses). In the shown range of detunings, i.e. within \SI{\pm100}{\giga\hertz}, the relationship between frequency and lifetime can be approximated by a straight line as \(\tau'\propto1/F_{P, Y_1\rightarrow Z_1}\) and \(F_{P, Y_1\rightarrow Z_1}\propto1/\nu\).}
	\label{fig:magnetic_tuning_supplement}
\end{figure}

To predict the effect of the photonic crystal waveguide (PCW) on the branching and lifetime of erbium dopants, a different model is required to predict the effective Purcell enhancement and thus the expected lifetimes, since erbium dopants in silicon have more than one radiative transition. To this end, we assume that in our experiments a \(4f\)-electron is resonantly excited to the lowest level of the excited state crystal-field (CF) manifold \(Y_1\) and can decay back to one of eight ground state CF levels \(Z_1...Z_8\) at a rate $A_{Z_i}$. The bulk lifetime $\tau$ is related to the decay constants of the individual transitions via
\begin{equation}
	1/\tau = A_{\mathrm{total}}=\sum_{i=1}^{8}A_{Z_i}.
\end{equation}
In our previous experiments~\cite{gritsch_narrow_2022}, the optical lifetime of the transition in a bulk silicon host was determined to be \SI{142\pm1}{\micro\second}, and the fraction of light emitted on the $Y_1 \rightarrow Z_1$ transition was $p_{Z_1}=\SI{23\pm5}{\percent}$. Recording the full emission spectrum in the decay of the $Y_1$ state by scanning a narrow-band filter now allows determining all individual decay rates:
\begin{equation}
	A_{Z_i} = A_{\mathrm{total}} \cdot p_{Z_i}.
\end{equation}
The resulting lifetime of dopants in the PCW is then given by:
\begin{equation}
	\tau'_{\mathrm{total}} = 1/(A_{Z_i}\cdot F_{P, Z_i}),
\end{equation}
where \(F_{P, Z_i}\) are the Purcell factors for each transition. To predict how the lifetime changes with frequency (gray curves in Fig.~4~b), we extract the Purcell factors \(F_{P, Y_1 \rightarrow Z_{2}}\) to \(F_{P, Y_1 \rightarrow Z_{8}}\) from the LDOS simulation and make the assumptions that they do not vary significantly between the different spatial positions/dipole orientations and also not for small changes in frequency \SI{<\pm100}{\giga\hertz}. The Purcell factor \(F_{P, Y_1 \rightarrow Z_{1}}\) affecting the lowest-to-lowest transition on the other hand, is computed using Eq.~\ref{eq:w1purcell} based on the measured \(n_g\) of three different waveguides, shown in~\ref{fig:supplementng} b. The measured lifetimes and the prediction based on the measured \(n_g\) of three identically designed waveguides is shown in Fig.~\ref{fig:magnetic_tuning_supplement}.

The minimum lifetime, however, is only achieved for emitters with a matching dipole at the location of maximum LDOS (cf. Fig.~\ref{fig:supplementldos}). However, for Er:Si, the optical dipole moment and its orientation are not known. In addition, the emitters in the studied devices will not be located precisely at the maximum. To account for this, we multiply a factor of 0.217 to the decay on the enhanced transition, $F_{P, Z_1}$, which is a free factor chosen to give the best agreement of the center gray curve in Fig.~4~b of the main text.

\section{Additional fluorescence spectra}

\begin{figure}
	\centering
	\includegraphics[width=1\linewidth]{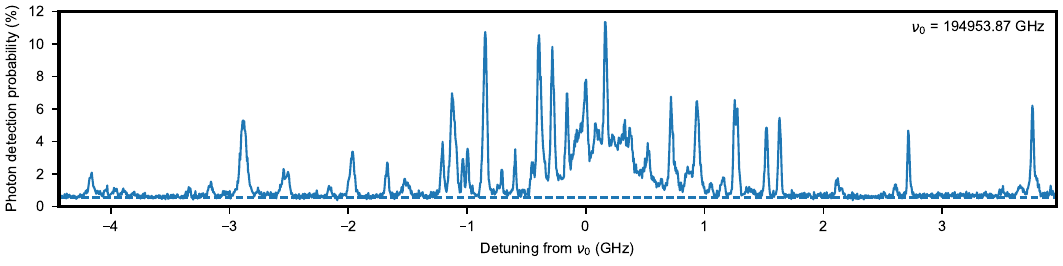}
	\caption{Fluorescence spectrum measured on waveguide B.}
	\label{fig:supplement_panorama_1003}
\end{figure}

As mentioned in the main text, we investigated three PCWs, A, B and C, with identical parameters except for their lengths. Fig.~3~a in the main text shows the fluorescence spectrum measured on PCW A. For completeness, Fig.~\ref{fig:supplement_panorama_1003} shows that of PCW B. Both are qualitatively and quantitatively very similar.

\section{Peak selection and fitting}

\begin{figure}
	\centering
	\includegraphics[width=1\linewidth]{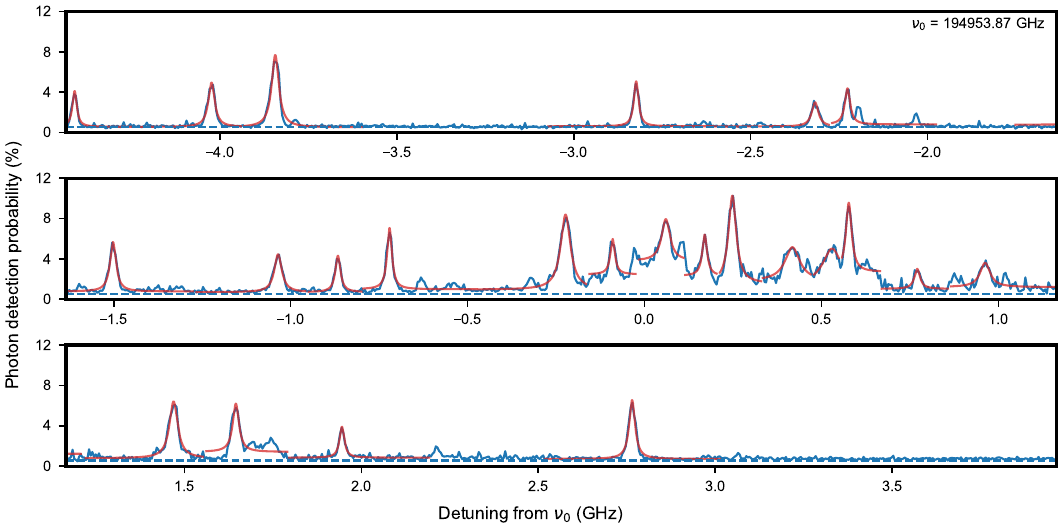}
	\caption{Fluorescence spectrum measured on waveguide B with automated Lorentzian fits.}
	\label{fig:supplement_linefits_1007}
\end{figure}

\begin{figure}
	\centering
	\includegraphics[width=1\linewidth]{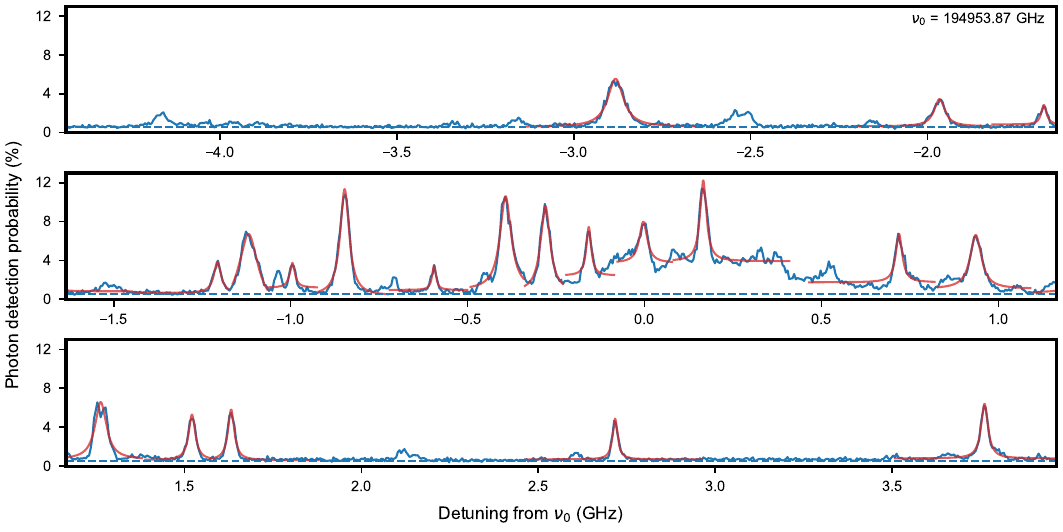}
	\caption{Fluorescence spectrum measured on waveguide A with automated Lorentzian fits.}
	\label{fig:supplement_linefits_1003}
\end{figure}

To extract the linewidths in Fig.~3~d of the main text, the relevant peaks in the fluorescence spectra are selected based on their prominence. We fit a Lorentzian emission line profile to all peaks with a prominence of at least 0.2. The resulting fits are shown in Fig.~\ref{fig:supplement_linefits_1007} for waveguide A and in Fig.~\ref{fig:supplement_linefits_1003} for waveguide B.

\section{Filter measurements}

\begin{figure}
	\centering
	\includegraphics[width=1\linewidth]{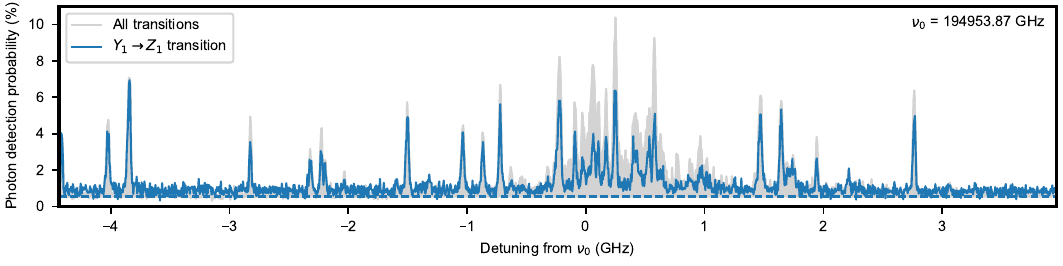}
	\caption{Fluorescence emission on the \(Y1\rightarrow Z1\) transition (blue) and emission on all (\(Y_1\rightarrow Z_1...z_8\)) transitions (gray) after resonant excitation to the \(Y_1\) excited state.}
	\label{fig:filtermeasurement}
\end{figure}

To determine the branching fraction of the \(Y_1\rightarrow Z_1\) transition, two separate fluorescence spectra are recorded. In both cases, the dopants are excited resonantly to the \(Y_1\) excited-state. Then, the fluorescence photons emitted on the \(Y_1\rightarrow Z_1..Z_8\) CF level transitions are once detected without discriminating by frequency and once with a narrow-band filter placed in the detection path that blocks photons emitted on the lower-frequency \(Y_1\rightarrow Z_2..Z_8\) transitions. The two spectra are used to determine the branching fraction in Figure~4~c of the main text; they are plotted in Fig.~\ref{fig:filtermeasurement} to allow for a comparison.

\section{Excitation efficiency}
A rigid upper-bound of the excitation efficiency is required to accurately investigate how the branching ratio between the $Z_1-Y_1$ transition and the suppressed $Z_i-Y_1$ transitions ($i \in \left[2,...,8\right]$) is changed in the PCW. This section discusses the details of how this estimate is obtained. In a first step, we determine the characteristic parameters for an emitter coupled to a strong driving field. For this, we measure Rabi oscillations as a function of pulse length on an ion in structure PCW C (see Methods) at frequency \SI{194954.17}{\giga\hertz}, which provides a good signal-to-noise ratio due to its brightness. 

\begin{figure}
	\centering
	\includegraphics[width=0.5\linewidth]{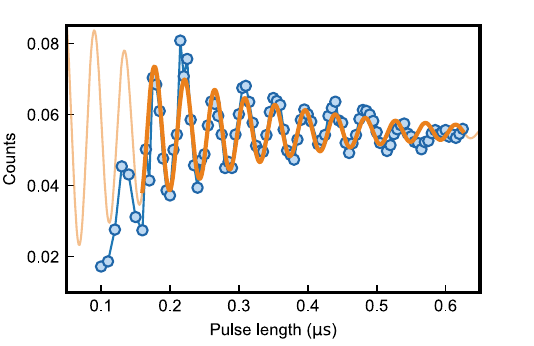}
	\caption{Pulsed resonant fluorescence with varying pulse length.  From the population oscillations observed when driving a single dopant, we extract a Rabi frequency of $\Omega=2\pi\times\SI{22.9(2)}{\mega\hertz}$ and dephasing $\gamma=\SI{10(1)}{\mega\hertz}$ using a least-square fit (solid line) of the function defined in eq.~\ref{eq:dampedrabi}. Data at pulse lengths below \SI{0.16}{\micro\second} is excluded from the fit because the pulse amplitude and phase are significantly influenced in this regime by the finite bandwidth of the acousto-optical modulator used for pulsing. An extension of the fit function beyond its range is plotted as a faint solid line.}     	\label{fig:rabi}
\end{figure}

To enable high-power driving, the electro-optic IQ-modulator is removed from the measurement setup for pulsed resonant fluorescence described in Appendix~C. We then apply laser pulses of varying duration and measure the subsequently emitted photons to determine the population of the driven ion. The recorded counts integrated over an interval of \SI{600}{\micro\second} after the pulse are plotted versus different pulse lengths in Fig.~\ref{fig:rabi}. 

The initial damping of the oscillation, observed for pulse lengths below \SI{0.15}{\micro\second}, is attributed to amplitude and phase changes caused by the finite bandwidth of the used acousto-optical modulators, which exhibit \SI{6}{\nano\second} and \SI{25}{\nano\second} rise times, respectively. Data in this range is excluded from the analysis.

From the data acquired at longer pulse durations, we extract the Rabi frequency by a least-squares fit (solid line in Fig.~\ref{fig:rabi}) to a damped Rabi oscillation of the form
\begin{equation}
    y(t) = \frac{a}{2} \left(1 - e^{-\gamma t/2} \cos(\Omega t+\phi)\right)+b,
    \label{eq:dampedrabi}
\end{equation}
where the parameter $a$ rescales to the measurement signal, $b$ accounts for background and detector dark counts, $\gamma$ is the dephasing, $\phi$ is a phase shift caused by the finite bandwidth of the pulse shapes, and $\Omega$ is the Rabi frequency. The relevant fit parameters are the Rabi frequency $\Omega =2 \pi \times \SI{22.9(2)}{\mega\hertz}$ and the dephasing $\gamma=\SI{10(1)}{\mega\hertz}$. 

By quantifying the insertion loss of the electro-optic IQ modulator used to obtain the data in Fig.~4 of the main text, and using the square-root scaling of the Rabi frequency with power, we can upper bound the achievable Rabi frequency in this measurement to $\Omega_\mathrm{max}\lesssim\SI{86}{\mega\hertz}$. We have compared the Rabi oscillations to those of another dopant on the same structure, which shows a shorter lifetime (\SI{152(7)}{\micro\second} compared to \SI{236(7)}{\micro\second}) but a lower number of collected photons. The extracted Rabi frequency is \SI{30}{\percent} lower, indicating the initially considered dopant is well coupled to the driving light field.

The extracted parameters are used to quantify the excitation efficiency of individual dopants in the measurements presented in Sec.~IID of the main text. To this end, we model the system using the optical Bloch equations, treating a two-level emitter with density matrix $\rho$ coupled to a strong driving field. Including a damping term $\gamma$ to reflect the dephasing, we numerically solve the differential equation
\begin{equation}
\frac{d}{dt}
\begin{pmatrix}
\operatorname{Re}\left[\rho_{12}\right] \\
\operatorname{Im}\left[\rho_{12}\right]\\
\rho_{22}-\rho_{11} \\
\end{pmatrix}
=
\begin{pmatrix}
-\gamma & \delta(t) & 0 \\
\delta(t) & -\gamma & -\Omega(t) \\
0 & \Omega(t) & 0
\end{pmatrix}
\cdot 
\begin{pmatrix}
\operatorname{Re}\left[\rho_{12}\right] \\
\operatorname{Im}\left[\rho_{12}\right]\\
\rho_{22}-\rho_{11}
\end{pmatrix}.
\end{equation}

In this calculation, the pulse parameters --- its length of \SI{2}{\micro\second} and the \SI{10}{\mega\hertz} linear chirp of the frequency around the resonance --- are included via the temporal dependence of the Rabi frequency $\Omega(t)$ and the detuning $\delta(t)$. The spontaneous emission decay is two orders of magnitude slower than the relevant dynamics and is thus negligible. The initial condition is that the dopant is in the ground state $\boldsymbol{\rho_0}=(0,0,-1)^\mathrm{T}$.

\begin{figure}
	\centering
	\includegraphics[width=1\linewidth]{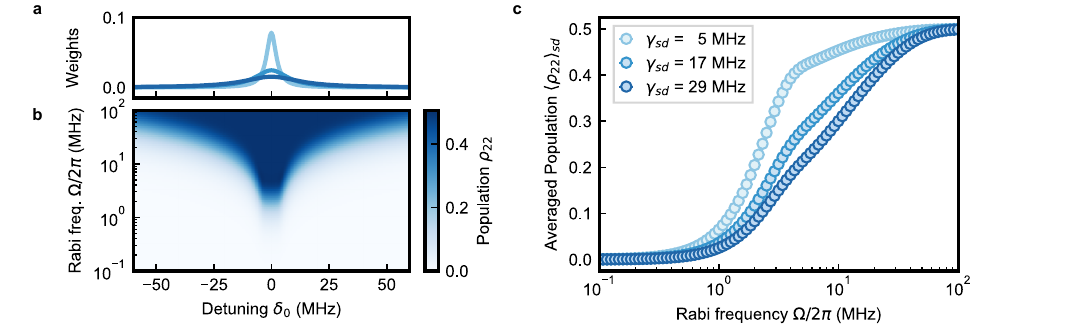}
	\caption{Simulation of the excitation efficiency upon optical driving. \textbf{a}, The dopants exhibit spectral diffusion with a Lorentzian distribution, such that the detuning between laser and dopant needs to be weighted with a corresponding factor when determining the average excitation probability. The shown distributions have the same color and FWHM as in the legend of panel c. \textbf{b}, Result of a simulation of the optical Bloch equations, showing the excited state population of a two-level emitter after a pulsed excitation of length \SI{2}{\micro\second} and a linear chirp of \SI{10}{\mega\hertz} of the driving field frequency around the center frequency. The simulation is performed for various Rabi frequencies $\Omega$ and fixed detunings $\delta_0$ from the center frequency. \textbf{c}, The averaged population of the excited state of an emitter subject to SD, calculated as a weighted average of values shown in panel a and b, is plotted for three spectral diffusion widths and varying driving field strengths. in the high-power limit, the population approaches, but never exceeds, $50\,\%$.}
	\label{fig:OBE_simulation}
\end{figure}

Because of the experimentally observed spectral diffusion, the resonant frequency of the dopants will differ from one repetition of the experiment to the next. To account for this, we solve the differential equation for a range of static offsets $\delta_0$ in the detuning parameter $\delta(t)$. The resulting excited state population is depicted in Fig.~\ref{fig:OBE_simulation}b. 
The dopant's spectral diffusion is then incorporated by averaging over the distribution of detunings. It can be estimated by the linewidth measurement $\Delta\nu$ of the ions, which is dominated by dephasing $\gamma$ and spectral diffusion $\gamma_{sd}$ such that $\Delta\nu\approx\gamma+\gamma_{sd}$. We simulate according to the measurements discussed in Sec.~IIB represented by the mean value $\Delta\nu_\mathrm{mean}=\SI{27}{\mega\hertz}$, a one standard deviation lower value $\Delta\nu_\mathrm{narrow}=\SI{15}{\mega\hertz}$, and a one standard deviation higher value $\Delta\nu_\mathrm{broad}=\SI{39}{\mega\hertz}$. The spectral diffusion follows a Lorentzian distribution and we calculate the ensemble average by 
\begin{equation}
    \langle \rho_{22}(\delta)\rangle_{sd} = \int \frac{1}{\pi\sigma_{sd}\left[1+\left(\frac{\delta}{\sigma_{sd}}\right)^2\right]}\rho_{22}(\delta) d\delta,  
\end{equation}
where $\sigma_{sd}=\gamma_{sd}/2$. We approximate the integral with a weighted sum and appropriate cutoffs as depicted in Fig.~\ref{fig:OBE_simulation}a. The resulting averaged population in the excited state for various Rabi frequencies is shown in Fig.~\ref{fig:OBE_simulation}c. For large Rabi frequencies, the population approaches, but never exceeds $50\,\%$ for all spectral diffusion parameters. Thus, independent of the precise laser power used, we can give an upper bound for the excitation efficiency of $\eta_{exc}=\SI{50}{\percent}$.

\section{Single-photon detection efficiency}

In this section, we rule out that the observed relative increase in \(Y_1\rightarrow Z_1\) emission is merely a result of spectral filtering by the PCW. To this end, we calculate an upper bound to the expected number of photons if the \(Y_1 \rightarrow Z_1\) branching fraction were unchanged compared to those previously determined for erbium dopants in bulk silicon, i.e. \(P_{Z1, \text{bulk}}=\SI{23\pm5}{\percent}\)~\cite{gritsch_narrow_2022}.

To this end, we first determine the probability that the emitter is excited after a laser pulse \(\eta_{exc}\). The used pulses have a length of \SI{2}{\micro\second} and are thus spectrally more narrow than the spectral diffusion linewidth of \(\approx\SI{23}{\mega\hertz}\). Still, as we use excitation pulses with a high Rabi frequency, the dopant will also be excited if it is off-resonant with the drive. A numerical simulation based on the optical Bloch equations averaged over the spectral diffusion linewidth leads to \(\eta_{exc} \approx \SI{50}{\percent}\) when assuming fast dephasing, and even lower values for slow dephasing.

Next, to derive an upper bound, we assume that an erbium dopant in the excited state only decays radiatively, and that all emitted photons are coupled into the PCW guided mode with a probability \(\beta \approx 1\)~\cite{javadi_numerical_2018}.

Finally, we calculate the probability that this photon will eventually be detected after being coupled out of the PCW and transmitted through the optical components of the setup. To this end, we perform an FDTD simulation with MEEP~\cite{oskooi_meep_2010} to determine the transmission efficiency of the step coupler and the PCW to strip waveguide interface \(\eta_{sc}=\SI{92\pm2}{\percent}\). Furthermore, we measure the maximum reflectivity \(R\) of the chips and perform an FDTD simulation to obtain the expected maximum reflectivity of the strip waveguide/step coupler interface \(R_{PCW} = \SI{85}{\percent}\), which allows determining the efficiency of fiber-to-chip (here chip-to-fiber) coupling of \(\eta_{ftc} = \sqrt{R} = \SI{73\pm4}{\percent}\). In addition, we calculate the losses in the detection path: The photon is transmitted through an SMF-28 ClearCurve to standard SMF-28 splice and through the cryostat fiber feed-through with a combined probability of \(\eta_{ft} = \SI{90\pm1}{\percent}\), through a 95:5 beam splitter with a probability of \(\eta_{bs} = \SI{95.0\pm1.5}{\percent}\), and through an optical switch with a probability of \(\eta_{os} = \SI{78\pm1}{\percent}\). If it was emitted on the \(Y_1\rightarrow Z1\) transition, it then passes through a narrow-band filter with a probability of \(\eta_{flt} = \SI{47.6\pm0.5}{\percent}\). Finally, the transmitted photons are detected by the SNSPD with a quantum efficiency of \(\eta_{qe} = \SI{75\pm5}{\percent}\).

If the branching fraction in the PCW were unchanged from the bulk value, we would thus expect to detect a photon emitted on the \(Y_1\rightarrow Z_1\) transition with a probability of at most:

\begin{equation}
    \begin{split}
        P &= \eta_{exc}\cdot P_{Z1, \text{bulk}}\cdot \beta \cdot \eta_{sc} \cdot \eta_{ftc}\cdot \eta_{ft}\cdot \eta_{bs}\cdot \eta_{os}\cdot \eta_{flt}\cdot \eta_{qe}\\ &= \SI{1.84\pm0.43}{\percent}.
    \end{split}
\end{equation}

This is an upper bound as we do not account for propagation losses in the PCW and emission into free space ( \(\beta < 1 \) ).

The expected experimental value of \SI{3.41\pm0.23}{\percent} is clearly above the expected value calculated above. This proves that the increase in relative emission on the \(Y_1 \rightarrow Z_1\) transition cannot be explained with spectral filtering alone, and instead we observe the effect of the spectrally selective inhibition of radiative decay channels.


\begin{thebibliography}{10}
\expandafter\ifx\csname url\endcsname\relax
  \def\url#1{\texttt{#1}}\fi
\expandafter\ifx\csname urlprefix\endcsname\relax\def\urlprefix{URL }\fi
\providecommand{\bibinfo}[2]{#2}
\providecommand{\eprint}[2][]{\url{#2}}

\bibitem{aharonovich_solid-state_2016}
\bibinfo{author}{Aharonovich, I.}, \bibinfo{author}{Englund, D.} \&
  \bibinfo{author}{Toth, M.}
\newblock \bibinfo{title}{Solid-state single-photon emitters}.
\newblock \emph{\bibinfo{journal}{Nature Photon.}}
  \textbf{\bibinfo{volume}{10}}, \bibinfo{pages}{631--641}
  (\bibinfo{year}{2016}).

\bibitem{reiserer_colloquium_2022}
\bibinfo{author}{Reiserer, A.}
\newblock \bibinfo{title}{Colloquium: {Cavity}-enhanced quantum network nodes}.
\newblock \emph{\bibinfo{journal}{Rev. Mod. Phys.}}
  \textbf{\bibinfo{volume}{94}}, \bibinfo{pages}{041003}
  (\bibinfo{year}{2022}).

\bibitem{chen_parallel_2020}
\bibinfo{author}{Chen, S.}, \bibinfo{author}{Raha, M.},
  \bibinfo{author}{Phenicie, C.~M.}, \bibinfo{author}{Ourari, S.} \&
  \bibinfo{author}{Thompson, J.~D.}
\newblock \bibinfo{title}{Parallel single-shot measurement and coherent control
  of solid-state spins below the diffraction limit}.
\newblock \emph{\bibinfo{journal}{Science}} \textbf{\bibinfo{volume}{370}},
  \bibinfo{pages}{592--595} (\bibinfo{year}{2020}).

\bibitem{fischer_signatures_2017}
\bibinfo{author}{Fischer, K.~A.} \emph{et~al.}
\newblock \bibinfo{title}{Signatures of two-photon pulses from a quantum
  two-level system}.
\newblock \emph{\bibinfo{journal}{Nature Physics}}
  \textbf{\bibinfo{volume}{13}}, \bibinfo{pages}{649--654}
  (\bibinfo{year}{2017}).

\bibitem{kleppner_inhibited_1981}
\bibinfo{author}{Kleppner, D.}
\newblock \bibinfo{title}{Inhibited {Spontaneous} {Emission}}.
\newblock \emph{\bibinfo{journal}{Phys. Rev. Lett.}}
  \textbf{\bibinfo{volume}{47}}, \bibinfo{pages}{233--236}
  (\bibinfo{year}{1981}).

\bibitem{yablonovitch_inhibited_1987}
\bibinfo{author}{Yablonovitch, E.}
\newblock \bibinfo{title}{Inhibited {Spontaneous} {Emission} in {Solid}-{State}
  {Physics} and {Electronics}}.
\newblock \emph{\bibinfo{journal}{Phys. Rev. Lett.}}
  \textbf{\bibinfo{volume}{58}}, \bibinfo{pages}{2059--2062}
  (\bibinfo{year}{1987}).

\bibitem{polman_erbium_2001}
\bibinfo{author}{Polman, A.}
\newblock \bibinfo{title}{Erbium as a probe of everything?}
\newblock \emph{\bibinfo{journal}{Physica B: Condensed Matter}}
  \textbf{\bibinfo{volume}{300}}, \bibinfo{pages}{78--90}
  (\bibinfo{year}{2001}).

\bibitem{zhao_suppression_2012}
\bibinfo{author}{Zhao, H.-Q.}, \bibinfo{author}{Fujiwara, M.} \&
  \bibinfo{author}{Takeuchi, S.}
\newblock \bibinfo{title}{Suppression of fluorescence phonon sideband from
  nitrogen vacancy centers in diamond nanocrystals by substrate effect}.
\newblock \emph{\bibinfo{journal}{Opt. Express}} \textbf{\bibinfo{volume}{20}},
  \bibinfo{pages}{15628--15635} (\bibinfo{year}{2012}).

\bibitem{karaveli_spectral_2011}
\bibinfo{author}{Karaveli, S.} \& \bibinfo{author}{Zia, R.}
\newblock \bibinfo{title}{Spectral {Tuning} by {Selective} {Enhancement} of
  {Electric} and {Magnetic} {Dipole} {Emission}}.
\newblock \emph{\bibinfo{journal}{Phys. Rev. Lett.}}
  \textbf{\bibinfo{volume}{106}}, \bibinfo{pages}{193004}
  (\bibinfo{year}{2011}).

\bibitem{lodahl_interfacing_2015}
\bibinfo{author}{Lodahl, P.}, \bibinfo{author}{Mahmoodian, S.} \&
  \bibinfo{author}{Stobbe, S.}
\newblock \bibinfo{title}{Interfacing single photons and single quantum dots
  with photonic nanostructures}.
\newblock \emph{\bibinfo{journal}{Rev. Mod. Phys.}}
  \textbf{\bibinfo{volume}{87}}, \bibinfo{pages}{347--400}
  (\bibinfo{year}{2015}).

\bibitem{bleuse_inhibition_2011}
\bibinfo{author}{Bleuse, J.} \emph{et~al.}
\newblock \bibinfo{title}{Inhibition, {Enhancement}, and {Control} of
  {Spontaneous} {Emission} in {Photonic} {Nanowires}}.
\newblock \emph{\bibinfo{journal}{Phys. Rev. Lett.}}
  \textbf{\bibinfo{volume}{106}}, \bibinfo{pages}{103601}
  (\bibinfo{year}{2011}).

\bibitem{viasnoff-schwoob_spontaneous_2005}
\bibinfo{author}{Viasnoff-Schwoob, E.} \emph{et~al.}
\newblock \bibinfo{title}{Spontaneous {Emission} {Enhancement} of {Quantum}
  {Dots} in a {Photonic} {Crystal} {Wire}}.
\newblock \emph{\bibinfo{journal}{Phys. Rev. Lett.}}
  \textbf{\bibinfo{volume}{95}}, \bibinfo{pages}{183901}
  (\bibinfo{year}{2005}).

\bibitem{wang_mapping_2011}
\bibinfo{author}{Wang, Q.}, \bibinfo{author}{Stobbe, S.} \&
  \bibinfo{author}{Lodahl, P.}
\newblock \bibinfo{title}{Mapping the {{Local Density}} of {{Optical States}}
  of a {{Photonic Crystal}} with {{Single Quantum Dots}}}.
\newblock \emph{\bibinfo{journal}{Physical Review Letters}}
  \textbf{\bibinfo{volume}{107}}, \bibinfo{pages}{167404}
  (\bibinfo{year}{2011}).

\bibitem{arcari_near-unity_2014}
\bibinfo{author}{Arcari, M.} \emph{et~al.}
\newblock \bibinfo{title}{Near-{Unity} {Coupling} {Efficiency} of a {Quantum}
  {Emitter} to a {Photonic} {Crystal} {Waveguide}}.
\newblock \emph{\bibinfo{journal}{Phys. Rev. Lett.}}
  \textbf{\bibinfo{volume}{113}}, \bibinfo{pages}{093603}
  (\bibinfo{year}{2014}).

\bibitem{ding_purcell-enhanced_2025}
\bibinfo{author}{Ding, S.~W.} \emph{et~al.}
\newblock \bibinfo{title}{Purcell-{Enhanced} {Emissions} from {Diamond} {Color}
  {Centers} in {Slow} {Light} {Photonic} {Crystal} {Waveguides}}.
\newblock \emph{\bibinfo{journal}{Nano Lett.}} \textbf{\bibinfo{volume}{25}},
  \bibinfo{pages}{12125--12131} (\bibinfo{year}{2025}).

\bibitem{gritsch_purcell_2023}
\bibinfo{author}{Gritsch, A.}, \bibinfo{author}{Ulanowski, A.} \&
  \bibinfo{author}{Reiserer, A.}
\newblock \bibinfo{title}{Purcell enhancement of single-photon emitters in
  silicon}.
\newblock \emph{\bibinfo{journal}{Optica}} \textbf{\bibinfo{volume}{10}},
  \bibinfo{pages}{783--789} (\bibinfo{year}{2023}).

\bibitem{fruh_spectral_2026}
\bibinfo{author}{Fr{\"u}h, J.}, \bibinfo{author}{Salamon, F.},
  \bibinfo{author}{Gritsch, A.}, \bibinfo{author}{Ulanowski, A.} \&
  \bibinfo{author}{Reiserer, A.}
\newblock \bibinfo{title}{Spectral {{Stability}} of {{Cavity-Enhanced
  Single-Photon Emitters}} in {{Silicon}}}.
\newblock \emph{\bibinfo{journal}{PRX Quantum}} \textbf{\bibinfo{volume}{7}},
  \bibinfo{pages}{020363} (\bibinfo{year}{2026}).

\bibitem{weiss_erbium_2021}
\bibinfo{author}{Weiss, L.}, \bibinfo{author}{Gritsch, A.},
  \bibinfo{author}{Merkel, B.} \& \bibinfo{author}{Reiserer, A.}
\newblock \bibinfo{title}{Erbium dopants in nanophotonic silicon waveguides}.
\newblock \emph{\bibinfo{journal}{Optica}} \textbf{\bibinfo{volume}{8}},
  \bibinfo{pages}{40--41} (\bibinfo{year}{2021}).

\bibitem{gritsch_narrow_2022}
\bibinfo{author}{Gritsch, A.}, \bibinfo{author}{Weiss, L.},
  \bibinfo{author}{Früh, J.}, \bibinfo{author}{Rinner, S.} \&
  \bibinfo{author}{Reiserer, A.}
\newblock \bibinfo{title}{Narrow {Optical} {Transitions} in
  {Erbium}-{Implanted} {Silicon} {Waveguides}}.
\newblock \emph{\bibinfo{journal}{Phys. Rev. X}} \textbf{\bibinfo{volume}{12}},
  \bibinfo{pages}{041009} (\bibinfo{year}{2022}).

\bibitem{berkman_observing_2023}
\bibinfo{author}{Berkman, I.~R.} \emph{et~al.}
\newblock \bibinfo{title}{Observing {Er3}+ {Sites} in {Si} {With} an {In}
  {Situ} {Single}-{Photon} {Detector}}.
\newblock \emph{\bibinfo{journal}{Phys. Rev. Appl.}}
  \textbf{\bibinfo{volume}{19}}, \bibinfo{pages}{014037}
  (\bibinfo{year}{2023}).

\bibitem{holewa_solid-state_2025}
\bibinfo{author}{Holewa, P.} \emph{et~al.}
\newblock \bibinfo{title}{Solid-state single-photon sources operating in the
  telecom wavelength range}.
\newblock \emph{\bibinfo{journal}{Nanophotonics}}
  \textbf{\bibinfo{volume}{14}}, \bibinfo{pages}{1729--1774}
  (\bibinfo{year}{2025}).

\bibitem{berkman_long_2025}
\bibinfo{author}{Berkman, I.~R.} \emph{et~al.}
\newblock \bibinfo{title}{Long optical and electron spin coherence times for
  erbium ions in silicon}.
\newblock \emph{\bibinfo{journal}{npj Quantum Inf.}}
  \textbf{\bibinfo{volume}{11}}, \bibinfo{pages}{1--10} (\bibinfo{year}{2025}).

\bibitem{rinner_erbium_2023}
\bibinfo{author}{Rinner, S.}, \bibinfo{author}{Burger, F.},
  \bibinfo{author}{Gritsch, A.}, \bibinfo{author}{Schmitt, J.} \&
  \bibinfo{author}{Reiserer, A.}
\newblock \bibinfo{title}{Erbium emitters in commercially fabricated
  nanophotonic silicon waveguides}.
\newblock \emph{\bibinfo{journal}{Nanophotonics}}
  \textbf{\bibinfo{volume}{12}}, \bibinfo{pages}{3455--3462}
  (\bibinfo{year}{2023}).

\bibitem{gritsch_optical_2025}
\bibinfo{author}{Gritsch, A.}, \bibinfo{author}{Ulanowski, A.},
  \bibinfo{author}{Pforr, J.} \& \bibinfo{author}{Reiserer, A.}
\newblock \bibinfo{title}{Optical single-shot readout of spin qubits in
  silicon}.
\newblock \emph{\bibinfo{journal}{Nat. Commun.}} \textbf{\bibinfo{volume}{16}},
  \bibinfo{pages}{64} (\bibinfo{year}{2025}).

\bibitem{holzapfel_characterization_2025}
\bibinfo{author}{Holzäpfel, A.} \emph{et~al.}
\newblock \bibinfo{title}{Characterization of the {Spin} and {Crystal} {Field}
  {Hamiltonian} of {Erbium} {Dopants} in {Silicon}}.
\newblock \emph{\bibinfo{journal}{Advanced Quantum Technologies}}
  \textbf{\bibinfo{volume}{8}}, \bibinfo{pages}{2400342}
  (\bibinfo{year}{2025}).

\bibitem{joannopoulos_photonic_2011}
\bibinfo{author}{Joannopoulos, J.~D.}, \bibinfo{author}{Johnson, S.~G.},
  \bibinfo{author}{Winn, J.~N.} \& \bibinfo{author}{Meade, R.~D.}
\newblock \emph{\bibinfo{title}{Photonic {Crystals}: {Molding} the {Flow} of
  {Light} - {Second} {Edition}}} (\bibinfo{publisher}{Princeton University
  Press}, \bibinfo{year}{2011}).

\bibitem{johnson_block-iterative_2001}
\bibinfo{author}{Johnson, S.~G.} \& \bibinfo{author}{Joannopoulos, J.~D.}
\newblock \bibinfo{title}{Block-iterative frequency-domain methods for
  {Maxwell}’s equations in a planewave basis}.
\newblock \emph{\bibinfo{journal}{Opt. Express}} \textbf{\bibinfo{volume}{8}},
  \bibinfo{pages}{173--190} (\bibinfo{year}{2001}).

\bibitem{vlasov_active_2005}
\bibinfo{author}{Vlasov, Y.~A.}, \bibinfo{author}{O'Boyle, M.},
  \bibinfo{author}{Hamann, H.~F.} \& \bibinfo{author}{McNab, S.~J.}
\newblock \bibinfo{title}{Active control of slow light on a chip with photonic
  crystal waveguides}.
\newblock \emph{\bibinfo{journal}{Nature}} \textbf{\bibinfo{volume}{438}},
  \bibinfo{pages}{65--69} (\bibinfo{year}{2005}).

\bibitem{baba_slow_2008}
\bibinfo{author}{Baba, T.}
\newblock \bibinfo{title}{Slow light in photonic crystals}.
\newblock \emph{\bibinfo{journal}{Nature Photon.}}
  \textbf{\bibinfo{volume}{2}}, \bibinfo{pages}{465--473}
  (\bibinfo{year}{2008}).

\bibitem{oskooi_meep_2010}
\bibinfo{author}{Oskooi, A.~F.} \emph{et~al.}
\newblock \bibinfo{title}{Meep: {A} flexible free-software package for
  electromagnetic simulations by the {FDTD} method}.
\newblock \emph{\bibinfo{journal}{Comput. Phys. Commun.}}
  \textbf{\bibinfo{volume}{181}}, \bibinfo{pages}{687--702}
  (\bibinfo{year}{2010}).

\bibitem{javadi_numerical_2018}
\bibinfo{author}{Javadi, A.}, \bibinfo{author}{Mahmoodian, S.},
  \bibinfo{author}{Söllner, I.} \& \bibinfo{author}{Lodahl, P.}
\newblock \bibinfo{title}{Numerical modeling of the coupling efficiency of
  single quantum emitters in photonic-crystal waveguides}.
\newblock \emph{\bibinfo{journal}{J. Opt. Soc. Am. B}}
  \textbf{\bibinfo{volume}{35}}, \bibinfo{pages}{514--522}
  (\bibinfo{year}{2018}).

\bibitem{gonzalez-tudela_lightmatter_2024}
\bibinfo{author}{González-Tudela, A.}, \bibinfo{author}{Reiserer, A.},
  \bibinfo{author}{García-Ripoll, J.~J.} \& \bibinfo{author}{García-Vidal,
  F.~J.}
\newblock \bibinfo{title}{Light–matter interactions in quantum nanophotonic
  devices}.
\newblock \emph{\bibinfo{journal}{Nat. Rev. Phys.}}
  \textbf{\bibinfo{volume}{6}}, \bibinfo{pages}{166--179}
  (\bibinfo{year}{2024}).

\bibitem{faggiani_implementing_2017}
\bibinfo{author}{Faggiani, R.}, \bibinfo{author}{Yang, J.},
  \bibinfo{author}{Hostein, R.} \& \bibinfo{author}{Lalanne, P.}
\newblock \bibinfo{title}{Implementing structural slow light on short length
  scales: the photonic speed bump}.
\newblock \emph{\bibinfo{journal}{Optica}} \textbf{\bibinfo{volume}{4}},
  \bibinfo{pages}{393--399} (\bibinfo{year}{2017}).

\bibitem{hughes_extrinsic_2005}
\bibinfo{author}{Hughes, S.}, \bibinfo{author}{Ramunno, L.},
  \bibinfo{author}{Young, J.~F.} \& \bibinfo{author}{Sipe, J.~E.}
\newblock \bibinfo{title}{Extrinsic {Optical} {Scattering} {Loss} in {Photonic}
  {Crystal} {Waveguides}: {Role} of {Fabrication} {Disorder} and {Photon}
  {Group} {Velocity}}.
\newblock \emph{\bibinfo{journal}{Phys. Rev. Lett.}}
  \textbf{\bibinfo{volume}{94}}, \bibinfo{pages}{033903}
  (\bibinfo{year}{2005}).

\bibitem{wolfowicz_quantum_2021}
\bibinfo{author}{Wolfowicz, G.} \emph{et~al.}
\newblock \bibinfo{title}{Quantum guidelines for solid-state spin defects}.
\newblock \emph{\bibinfo{journal}{Nat. Rev. Mater.}}
  \textbf{\bibinfo{volume}{6}}, \bibinfo{pages}{906} (\bibinfo{year}{2021}).

\bibitem{uysal_spin-photon_2025}
\bibinfo{author}{Uysal, M.~T.} \emph{et~al.}
\newblock \bibinfo{title}{Spin-{Photon} {Entanglement} of a {Single} {Er} {Ion}
  in the {Telecom} {Band}}.
\newblock \emph{\bibinfo{journal}{Phys. Rev. X}} \textbf{\bibinfo{volume}{15}},
  \bibinfo{pages}{011071} (\bibinfo{year}{2025}).

\bibitem{ruskuc_multiplexed_2025}
\bibinfo{author}{Ruskuc, A.} \emph{et~al.}
\newblock \bibinfo{title}{Multiplexed entanglement of multi-emitter quantum
  network nodes}.
\newblock \emph{\bibinfo{journal}{Nature}} \textbf{\bibinfo{volume}{639}},
  \bibinfo{pages}{54--59} (\bibinfo{year}{2025}).

\bibitem{craiciu_multifunctional_2021}
\bibinfo{author}{Craiciu, I.}, \bibinfo{author}{Lei, M.},
  \bibinfo{author}{Rochman, J.}, \bibinfo{author}{Bartholomew, J.~G.} \&
  \bibinfo{author}{Faraon, A.}
\newblock \bibinfo{title}{Multifunctional on-chip storage at telecommunication
  wavelength for quantum networks}.
\newblock \emph{\bibinfo{journal}{Optica}} \textbf{\bibinfo{volume}{8}},
  \bibinfo{pages}{114--121} (\bibinfo{year}{2021}).

\bibitem{tiranov_sub-second_2026}
\bibinfo{author}{Tiranov, A.} \emph{et~al.}
\newblock \bibinfo{title}{Sub-second spin and lifetime-limited optical
  coherences in {{171Yb3}}+:{{CaWO4}}}.
\newblock \emph{\bibinfo{journal}{Nature Communications}}
  \textbf{\bibinfo{volume}{17}}, \bibinfo{pages}{4115} (\bibinfo{year}{2026}).

\bibitem{raha_optical_2020}
\bibinfo{author}{Raha, M.} \emph{et~al.}
\newblock \bibinfo{title}{Optical quantum nondemolition measurement of a single
  rare earth ion qubit}.
\newblock \emph{\bibinfo{journal}{Nat. Commun.}} \textbf{\bibinfo{volume}{11}},
  \bibinfo{pages}{1605} (\bibinfo{year}{2020}).

\bibitem{lodahl_chiral_2017}
\bibinfo{author}{Lodahl, P.} \emph{et~al.}
\newblock \bibinfo{title}{Chiral quantum optics}.
\newblock \emph{\bibinfo{journal}{Nature}} \textbf{\bibinfo{volume}{541}},
  \bibinfo{pages}{473--480} (\bibinfo{year}{2017}).

\bibitem{uppu_quantum-dot-based_2021}
\bibinfo{author}{Uppu, R.}, \bibinfo{author}{Midolo, L.},
  \bibinfo{author}{Zhou, X.}, \bibinfo{author}{Carolan, J.} \&
  \bibinfo{author}{Lodahl, P.}
\newblock \bibinfo{title}{Quantum-dot-based deterministic photon–emitter
  interfaces for scalable photonic quantum technology}.
\newblock \emph{\bibinfo{journal}{Nat. Nanotechnol.}}
  \textbf{\bibinfo{volume}{16}}, \bibinfo{pages}{1308--1317}
  (\bibinfo{year}{2021}).

\bibitem{anderson_electrical_2019}
\bibinfo{author}{Anderson, C.~P.} \emph{et~al.}
\newblock \bibinfo{title}{Electrical and optical control of single spins
  integrated in scalable semiconductor devices}.
\newblock \emph{\bibinfo{journal}{Science}} \textbf{\bibinfo{volume}{366}},
  \bibinfo{pages}{1225--1230} (\bibinfo{year}{2019}).

\bibitem{hollenbach_wafer-scale_2022}
\bibinfo{author}{Hollenbach, M.} \emph{et~al.}
\newblock \bibinfo{title}{Wafer-scale nanofabrication of telecom single-photon
  emitters in silicon}.
\newblock \emph{\bibinfo{journal}{Nat. Commun.}} \textbf{\bibinfo{volume}{13}},
  \bibinfo{pages}{7683} (\bibinfo{year}{2022}).

\bibitem{tiranov_collective_2023}
\bibinfo{author}{Tiranov, A.} \emph{et~al.}
\newblock \bibinfo{title}{Collective super- and subradiant dynamics between
  distant optical quantum emitters}.
\newblock \emph{\bibinfo{journal}{Science}} \textbf{\bibinfo{volume}{379}},
  \bibinfo{pages}{389--393} (\bibinfo{year}{2023}).

\bibitem{ourari_indistinguishable_2023}
\bibinfo{author}{Ourari, S.} \emph{et~al.}
\newblock \bibinfo{title}{Indistinguishable telecom band photons from a single
  {Er} ion in the solid state}.
\newblock \emph{\bibinfo{journal}{Nature}} \textbf{\bibinfo{volume}{620}},
  \bibinfo{pages}{977--981} (\bibinfo{year}{2023}).

\bibitem{durand_broad_2021}
\bibinfo{author}{Durand, A.} \emph{et~al.}
\newblock \bibinfo{title}{Broad {Diversity} of {Near}-{Infrared}
  {Single}-{Photon} {Emitters} in {Silicon}}.
\newblock \emph{\bibinfo{journal}{Phys. Rev. Lett.}}
  \textbf{\bibinfo{volume}{126}}, \bibinfo{pages}{083602}
  (\bibinfo{year}{2021}).

\bibitem{higginbottom_optical_2022}
\bibinfo{author}{Higginbottom, D.~B.} \emph{et~al.}
\newblock \bibinfo{title}{Optical observation of single spins in silicon}.
\newblock \emph{\bibinfo{journal}{Nature}} \textbf{\bibinfo{volume}{607}},
  \bibinfo{pages}{266--270} (\bibinfo{year}{2022}).

\bibitem{komza_indistinguishable_2024}
\bibinfo{author}{Komza, L.} \emph{et~al.}
\newblock \bibinfo{title}{Indistinguishable photons from an artificial atom in
  silicon photonics}.
\newblock \emph{\bibinfo{journal}{Nat. Commun.}} \textbf{\bibinfo{volume}{15}},
  \bibinfo{pages}{6920} (\bibinfo{year}{2024}).

\bibitem{johnston_cavity-coupled_2024}
\bibinfo{author}{Johnston, A.}, \bibinfo{author}{Felix-Rendon, U.},
  \bibinfo{author}{Wong, Y.-E.} \& \bibinfo{author}{Chen, S.}
\newblock \bibinfo{title}{Cavity-coupled telecom atomic source in silicon}.
\newblock \emph{\bibinfo{journal}{Nat. Commun.}} \textbf{\bibinfo{volume}{15}},
  \bibinfo{pages}{2350} (\bibinfo{year}{2024}).

\bibitem{wolfowicz_vanadium_2020}
\bibinfo{author}{Wolfowicz, G.} \emph{et~al.}
\newblock \bibinfo{title}{Vanadium spin qubits as telecom quantum emitters in
  silicon carbide}.
\newblock \emph{\bibinfo{journal}{Sci. Adv.}} \textbf{\bibinfo{volume}{6}},
  \bibinfo{pages}{eaaz1192} (\bibinfo{year}{2020}).

\bibitem{lukin_integrated_2020}
\bibinfo{author}{Lukin, D.~M.}, \bibinfo{author}{Guidry, M.~A.} \&
  \bibinfo{author}{Vučković, J.}
\newblock \bibinfo{title}{Integrated {Quantum} {Photonics} with {Silicon}
  {Carbide}: {Challenges} and {Prospects}}.
\newblock \emph{\bibinfo{journal}{PRX Quantum}} \textbf{\bibinfo{volume}{1}},
  \bibinfo{pages}{020102} (\bibinfo{year}{2020}).

\bibitem{cilibrizzi_ultra-narrow_2023}
\bibinfo{author}{Cilibrizzi, P.} \emph{et~al.}
\newblock \bibinfo{title}{Ultra-narrow inhomogeneous spectral distribution of
  telecom-wavelength vanadium centres in isotopically-enriched silicon
  carbide}.
\newblock \emph{\bibinfo{journal}{Nat. Commun.}} \textbf{\bibinfo{volume}{14}},
  \bibinfo{pages}{8448} (\bibinfo{year}{2023}).

\bibitem{wan_large-scale_2020}
\bibinfo{author}{Wan, N.~H.} \emph{et~al.}
\newblock \bibinfo{title}{Large-scale integration of artificial atoms in hybrid
  photonic circuits}.
\newblock \emph{\bibinfo{journal}{Nature}} \textbf{\bibinfo{volume}{583}},
  \bibinfo{pages}{226--231} (\bibinfo{year}{2020}).

\bibitem{jia_two-pattern_2011}
\bibinfo{author}{Jia, L.} \& \bibinfo{author}{Thomas, E.~L.}
\newblock \bibinfo{title}{Two-pattern compound photonic crystals with a large
  complete photonic band gap}.
\newblock \emph{\bibinfo{journal}{Phys. Rev. A}} \textbf{\bibinfo{volume}{84}},
  \bibinfo{pages}{033810} (\bibinfo{year}{2011}).

\bibitem{hensen_loophole-free_2015}
\bibinfo{author}{Hensen, B.} \emph{et~al.}
\newblock \bibinfo{title}{Loophole-free {Bell} inequality violation using
  electron spins separated by 1.3 kilometres}.
\newblock \emph{\bibinfo{journal}{Nature}} \textbf{\bibinfo{volume}{526}},
  \bibinfo{pages}{682--686} (\bibinfo{year}{2015}).

\bibitem{sapienza_cavity_2010}
\bibinfo{author}{Sapienza, L.} \emph{et~al.}
\newblock \bibinfo{title}{Cavity {Quantum} {Electrodynamics} with
  {Anderson}-{Localized} {Modes}}.
\newblock \emph{\bibinfo{journal}{Science}} \textbf{\bibinfo{volume}{327}},
  \bibinfo{pages}{1352--1355} (\bibinfo{year}{2010}).

\bibitem{hugonin_coupling_2007}
\bibinfo{author}{Hugonin, J.~P.}, \bibinfo{author}{Lalanne, P.},
  \bibinfo{author}{White, T.~P.} \& \bibinfo{author}{Krauss, T.~F.}
\newblock \bibinfo{title}{Coupling into slow-mode photonic crystal waveguides}.
\newblock \emph{\bibinfo{journal}{Opt. Lett.}} \textbf{\bibinfo{volume}{32}},
  \bibinfo{pages}{2638--2640} (\bibinfo{year}{2007}).

\bibitem{tiecke_efficient_2015}
\bibinfo{author}{Tiecke, T.~G.} \emph{et~al.}
\newblock \bibinfo{title}{Efficient fiber-optical interface for nanophotonic
  devices}.
\newblock \emph{\bibinfo{journal}{Optica}} \textbf{\bibinfo{volume}{2}},
  \bibinfo{pages}{70--75} (\bibinfo{year}{2015}).

\bibitem{becher_nonclassical_2001}
\bibinfo{author}{Becher, C.} \emph{et~al.}
\newblock \bibinfo{title}{Nonclassical radiation from a single self-assembled
  {InAs} quantum dot}.
\newblock \emph{\bibinfo{journal}{Phys. Rev. B}} \textbf{\bibinfo{volume}{63}},
  \bibinfo{pages}{121312} (\bibinfo{year}{2001}).

\end{thebibliography}

\begin{thebibliography}{11}
\expandafter\ifx\csname url\endcsname\relax
  \def\url#1{\texttt{#1}}\fi
\expandafter\ifx\csname urlprefix\endcsname\relax\def\urlprefix{URL }\fi
\providecommand{\bibinfo}[2]{#2}
\providecommand{\eprint}[2][]{\url{#2}}

\bibitem{oskooi_meep_2010}
\bibinfo{author}{Oskooi, A.~F.} \emph{et~al.}
\newblock \bibinfo{title}{Meep: {A} flexible free-software package for
  electromagnetic simulations by the {FDTD} method}.
\newblock \emph{\bibinfo{journal}{Comput. Phys. Commun.}}
  \textbf{\bibinfo{volume}{181}}, \bibinfo{pages}{687--702}
  (\bibinfo{year}{2010}).

\bibitem{faggiani_implementing_2017}
\bibinfo{author}{Faggiani, R.}, \bibinfo{author}{Yang, J.},
  \bibinfo{author}{Hostein, R.} \& \bibinfo{author}{Lalanne, P.}
\newblock \bibinfo{title}{Implementing structural slow light on short length
  scales: the photonic speed bump}.
\newblock \emph{\bibinfo{journal}{Optica}} \textbf{\bibinfo{volume}{4}},
  \bibinfo{pages}{393--399} (\bibinfo{year}{2017}).

\bibitem{lodahl_interfacing_2015}
\bibinfo{author}{Lodahl, P.}, \bibinfo{author}{Mahmoodian, S.} \&
  \bibinfo{author}{Stobbe, S.}
\newblock \bibinfo{title}{Interfacing single photons and single quantum dots
  with photonic nanostructures}.
\newblock \emph{\bibinfo{journal}{Rev. Mod. Phys.}}
  \textbf{\bibinfo{volume}{87}}, \bibinfo{pages}{347--400}
  (\bibinfo{year}{2015}).

\bibitem{gritsch_narrow_2022}
\bibinfo{author}{Gritsch, A.}, \bibinfo{author}{Weiss, L.},
  \bibinfo{author}{Früh, J.}, \bibinfo{author}{Rinner, S.} \&
  \bibinfo{author}{Reiserer, A.}
\newblock \bibinfo{title}{Narrow {Optical} {Transitions} in
  {Erbium}-{Implanted} {Silicon} {Waveguides}}.
\newblock \emph{\bibinfo{journal}{Phys. Rev. X}} \textbf{\bibinfo{volume}{12}},
  \bibinfo{pages}{041009} (\bibinfo{year}{2022}).

\bibitem{javadi_numerical_2018}
\bibinfo{author}{Javadi, A.}, \bibinfo{author}{Mahmoodian, S.},
  \bibinfo{author}{Söllner, I.} \& \bibinfo{author}{Lodahl, P.}
\newblock \bibinfo{title}{Numerical modeling of the coupling efficiency of
  single quantum emitters in photonic-crystal waveguides}.
\newblock \emph{\bibinfo{journal}{J. Opt. Soc. Am. B}}
  \textbf{\bibinfo{volume}{35}}, \bibinfo{pages}{514--522}
  (\bibinfo{year}{2018}).

\end{thebibliography}
\end{document}